\documentclass[11pt,a4paper]{article} 
\usepackage{jheppubori,hyperref} 
\usepackage{diagbox}
\usepackage{comment}
\usepackage{bigints}
\usepackage{ulem}
\usepackage{amsmath}
\usepackage{graphicx}
\usepackage{subfigure}
\usepackage{amssymb}
\usepackage{xcolor}
\usepackage{multirow}
\usepackage{cancel}
\usepackage{color}
\usepackage{mathtools}
\usepackage{listings}
\usepackage{xcolor}
\usepackage{float}
\usepackage{slashed}
\usepackage{bm}
\usepackage{amsmath}
\usepackage{wrapfig}
\usepackage[toc,title,page]{appendix}
\usepackage[colorinlistoftodos]{todonotes}
\usepackage{lineno}
\usepackage{marginnote}

\newcommand{\be}{\begin{equation}}
\newcommand{\ee}{\end{equation}}
\newcommand{\beq}{\begin{equation}}
\newcommand{\eeq}{\end{equation}}
\newcommand{\bea}{\begin{eqnarray}}
\newcommand{\eea}{\end{eqnarray}}
\definecolor{airforceblue}{rgb}{0.36, 0.54, 0.66}
\definecolor{steelblue}{rgb}{0.27, 0.51, 0.71}
\definecolor{amber}{rgb}{1.0, 0.49, 0.0}

\newcommand{\s}{\slashed}

\newcommand{\sloops}{{\tt SloopS}\;}
\newcommand{\msbar}{${\overline{\text{MS}}}\;$}

\title{\boldmath Higgs-strahlung at the LHC in the inert doublet model.}

\author[a]{Dazhuang He,}
\affiliation[a]{Institute of Theoretical Physics, School of Physics, Dalian University of Technology, \\ No.2 Linggong Road, Dalian, Liaoning, 116024, P.R.China }
\author[b]{Yu Zhang,}
\affiliation[b]{School of Physics, Hefei University of Technology, No.420 Feicui Road,\\ Hefei, Anhui, 230601, P.R.China}
\author[c]{Fawzi Boudjema,}
\affiliation[c]{LAPTh, Universit\'e Savoie Mont Blanc, CNRS, BP~110, F-74000 Annecy, France}
\author[a]{and Hao Sun}
\emailAdd{dzhe@mail.dlut.edu.cn}
\emailAdd{dayu@hfut.edu.cn}
\emailAdd{boudjema@lapth.cnrs.fr}
\emailAdd{haosun@dlut.edu.cn}

\abstract{
$pp \to W^{\pm} h, Zh$  processes at the LHC are studied in the framework of  the inert doublet model (IDM). To quantify the effects of the IDM and their observability in these processes we revisit the NLO (QCD and EW) predictions in the Standard Model (SM) and their uncertainty. Taking all available current constraints on the parameter space of the IDM, we consider both the case of the IDM providing a good Dark Matter (DM) candidate within the freeze-out mechanism as well as when the DM restrictions are relaxed.  In the former, deviations from the SM of only a few per mil in these cross sections at the LHC are found and may not be measured.
In the latter, the deviations can reach a few percents and should be observable. Smaller discrepancies in this case require that the theoretical uncertainties be improved, in particular those arising from the parton distribution functions (PDFs). We stress the importance of the photon-induced real corrections and the need for further improvement in the extraction of the photon PDF. The analysis also showcases the development and exploitation of our automated tool for the computation of one-loop electroweak and QCD corrections for a New Physics model with internal tests such as those concerning the soft and collinear parts provided through both dipole subtraction and phase space slicing besides tests for ultra-violet finiteness and gauge-parameter independence.  

\vspace{0.5cm}
}

\begin{document}
\maketitle

\section{Introduction}
\label{sec::intro}
The discovery of the Higgs boson \cite{ATLAS:2012yve,CMS:2012qbp} at the Large Hadron Collider (LHC) in 2012 was the crowning of the Standard Model (SM) and the validation of mass generation through the mechanism of electroweak symmetry breaking. There is some unease with the rather light mass of the Higgs boson, $M_h=125$ GeV,  compared to, for example, the large scales derived from the indirect effect of the New Physics in flavour observables. Yet, Higgs observables can now be used to probe the structure of electroweak symmetry breaking. One approach is a model-independent approach in the context of effective operators and their effects on a large set of observables, not necessarily Higgs related. The other approach is within a coherent Ultraviolet, UV, completion, {\it i.e.,} a fully fledged model. If the model is unitary and renormalisable, a programme based on precision calculations can be conducted. This programme includes direct searches for, in principle, the particle content of the model but also the indirect effects of these particles, and modified couplings, in standard model processes. One aim of this paper is to concentrate on the indirect effects. Considering that the SM has passed so many tests with flying colours, one can ask whether any extension is already so constrained through existing observables now that prospects for seeing any deviation as a sign of New Physics (NP) in the future, in particular at the LHC, are slim. To be able to return a sound answer, a better determination of the theoretical uncertainties in SM predictions is in order, as well as more precise experimental measurements. In an approach based on a well-defined model, it is best to first start these investigations with a rather simplified, UV-complete model. The Inert Doublet Model, IDM \cite{Deshpande:1977rw,Barbieri:2006dq,Hambye:2007vf}, is one such model. It is a simplified two-scalar doublet model (2HDM) containing only an extra scalar doublet to the SM content. 
The IDM has a rich phenomenology both at the colliders and in Dark Matter (DM) physics, and as such has been widely studied \cite{Kanemura:2002vm,LopezHonorez:2006gr,Cao:2007rm,Gustafsson:2007pc,Agrawal:2008xz,Lundstrom:2008ai,Hambye:2009pw,Andreas:2009hj, Arina:2009um, Dolle:2009ft,Nezri:2009jd, Ferreira:2009jb,Miao:2010rg,Gong:2012ri, Gustafsson:2012aj, Swiezewska:2012eh, Wang:2012zv,Arhrib:2012ia,Klasen:2013btp,   Aoki:2013lhm,Ho:2013spa,Goudelis:2013uca,Arhrib:2013ela,Krawczyk:2013jta,Osland:2013sla,Ginzburg:2014ora,Abe:2015rja,Arhrib:2015hoa,Blinov:2015qva, Diaz:2015pyv,Ilnicka:2015jba,Belanger:2015kga,Carmona:2015haa,Queiroz:2015utg,Ferreira:2015pfi,Garcia-Cely:2015khw,Hashemi:2015swh,Datta:2016nfz,Akeroyd:2016ymd,Kanemura:2016sos,Belyaev:2016lok,Banerjee:2016vrp,Poulose:2016lvz,Eiteneuer:2017hoh,Senaha:2018xek,Ilnicka:2018def,Kalinowski:2018ylg,Arcadi:2019lka, Braathen:2019pxr, Banerjee:2019luv,Bhardwaj:2019mts,Zarnecki:2019poj,Sokolowska:2019xhe, Basu:2020qoe,Abouabid:2020eik,Zarnecki:2020swm,Zarnecki:2020nnw,Kalinowski:2020rmb,Banerjee:2021oxc,Banerjee:2021anv,Banerjee:2021xdp,Banerjee:2021hal,Robens:2021rkl,Robens:2021zvr,Robens:2021yrl,Ramsey-Musolf:2021ldh,Ghosh:2021noq,Aiko:2021nkb,Aiko:2022jbp,Abouabid:2022rnd} mostly in tree-level analyses, but also including higher-order corrections, either electroweak (EW) or QCD.

In this paper we quantify the indirect impact of the IDM on the Higgs-strahlung, ($pp \to Vh, V=W^\pm,Z$) cross sections and distributions at the LHC once the current constraints on the model (theoretical considerations, precision electroweak observables, non-observation of the heavy scalars in direct collider searches, impact on Higgs decays,..) are imposed. If the IDM is to qualify as a model of DM, we study the additional constraints set by DM observables, in particular the relic density and direct detection. The indirect effects of the IDM in Higgs-strahlung processes necessarily involve one-loop corrections, which, for the IDM, take place in the electroweak sector. Nevertheless, the issue of the SM theoretical uncertainties arises, these are both of QCD and electroweak nature in the context of the LHC. 

From an experimental point of view, the production of $Vh$ in the dominant channel $h \to b \bar b$ has been observed by ATLAS and CMS \cite{ATLAS:2018kot,CMS:2018nsn} with a signal strength within approximately $25\%$ of the SM value. 
Evidence in the $h \to \tau \bar \tau$ channel has also been reported \cite{ATLAS:2023qpu}. 
In the future, more analyses and combinations will be performed. The {\it experimental} accuracy in the various channels will be improved.  
For example, the precision on the $b \bar b$ channel is expected to reach $15 \%$ ($5\%$) on the $Wh$~($Zh)~$\cite{ATLAS:2018jlh,CMS:2018qgz} in the high luminosity LHC (HL-LHC). 
We are confident that by the time the HL-LHC data are exploited, other $Vh$ channels will have been included and further improvements in the $h \to b \bar b$ signature will have been achieved. 
At the same time, the accuracy of the PDF sets will have improved through the exploitation of large LHC data sets. This will reduce an important part of the theoretical uncertainties.

The other aim of this paper is to showcase the improvements made in, {\tt SloopS}~\cite{Boudjema:2005hb,Baro:2007em,Baro:2008bg,Baro:2009na,Boudjema:2011ig,Boudjema:2014gza,Belanger:2016tqb,Belanger:2017rgu,Banerjee:2019luv}), our automatic one-loop calculator that now performs both QCD and electroweak corrections and that handles both the SM and a new physics model (the IDM in this case) beyond leading order. In this respect, we go through the different elements (QCD and EW)  that build up the SM and the IDM contributions to  Higgs-strahlung beyond tree-level.

The structure of this paper is as follows. In Section \ref{sec::Def_IDM} we briefly describe the Lagrangian of the IDM and underline the set of physical parameters necessary to define the model. By closely following the extensive study we performed in \cite{Banerjee:2021oxc}, which we urge the reader to consult, we update the allowed parameter space (theoretical considerations and current experimental constraints) in the parameter space of the model. We classify models within two categories: those in which the IDM does not provide a DM candidate but passes all other constraints, and those in which the IDM furnishes a DM candidate. Within each category, we propose some benchmark points. 

In Sect.~\ref{sec::vhsm}, the LO and NLO, QCD and electroweak calculations for Higgs-strahlung are discussed and their implementation in our code addressed. 
Since these LHC processes set an example for automation\footnote{Other NLO tools employing different techniques, with varying degrees of automation and range of application  have very recently been nicely reviewed in~\cite{Campbell:2022qmc} together with their interfacing with general-purpose Monte Carlo event generators and shower programmes.} of NLO processes in {\tt SloopS}, with both QCD and EW corrections within and beyond the SM,  we go through the generation of the virtual and different parts of the real (bremsstrahlung, $gq/\gamma q$ and $gg/\gamma \gamma$ induced) corrections. Tuned comparisons for the QCD NLO corrections with {\tt MadGraph} \cite{Alwall:2014hca} are performed and found to be excellent. For electroweak corrections we use an OS (on-shell scheme) but contrary to many of our previous studies
({\tt SloopS}~\cite{Boudjema:2005hb,Baro:2007em,Baro:2008bg,Baro:2009na,Boudjema:2011ig,Boudjema:2014gza,Belanger:2016tqb,Belanger:2017rgu,Banerjee:2019luv}) we use the so-called $\alpha_{G_\mu}$ scheme instead of the fine structure constant defined in the Thomson limit. This merges well with the QCD corrections and allows for the use of massless fermions. Also, new compared to our previous studies of NLO corrections in handling infrared/collinear divergences, we use both the dipole subtraction \cite{Catani:1996jh,Catani:1996vz,Catani:2002hc} and  the  two-cutoff phase space slicing methods \cite{Harris:2001sx}. This serves as an additional check for our calculations for both QCD and EW corrections. We underline the importance of the photon-induced corrections and the need for a good determination of the photon PDF. Sect.~\ref{sec::hv_idm} is devoted to the generation of one-loop additional corrections within the IDM.  Results for the IDM contribution on the inclusive cross sections for a large scan on the allowed parameter space are presented. We also study, for some benchmark points, the effects on the transverse momentum and rapidity of the Higgs. We wrap up with a concise summary in Section \ref{sec::summary}.

\section{The Inert Higgs Doublet Model, description and constraints}
\label{sec::Def_IDM}

\subsection{A brief description of the IDM}
Beside the SM scalar doublet $\Phi_1$, the IDM includes a second scalar doublet, $\Phi_2$, which does not take part in electroweak symmetry breaking. A discrete $\mathbb{Z}_2 $ symmetry is built in whereby all SM fields are even, such that under this symmetry transformation $\Phi_1 \to \Phi_1$, whereas the new scalars are odd $, \Phi_2 \to -\Phi_2$. This symmetry prevents the new scalars to couple to the fermions of the SM through renormalisable operators. It also restricts the form of the (renormalisable) Lagrangian, in particular the extended scalar potential.

Following the notations we gave in Ref.~\cite{Banerjee:2021oxc}, the IDM Lagrangian is 
\begin{equation}\label{LIDM}
\begin{aligned}
\mathcal{L}_{\rm IDM}^{\text{scalar}}& =(D^\mu \Phi_1)^\dagger D_\mu \Phi_1+ (D^\mu \Phi_2)^\dagger D_\mu \Phi_2 -V_{\mathrm{IDM}}(\Phi_{1}, \Phi_{2}), \\
\end{aligned}
\end{equation}
with the scalar potential
\begin{equation}\label{VIDM}
\begin{aligned}
V_{\mathrm{IDM}}(\Phi_{1}, \Phi_{2}) &=
\mu_{1}^{2}|\Phi_{1}|^{2}+\mu_{2}^{2}|\Phi_{2}|^{2} +\lambda_{1}|\Phi_{1}|^{4}+\lambda_{2}|\Phi_{2}|^{4}
+\lambda_{3}|\Phi_{1}|^{2}|\Phi_{2}|^{2}+\lambda_{4}|\Phi_{1}^{\dagger}\Phi_{2}|^{2}  \\
&\quad
+\frac{1}{2}\lambda_{5}\left[\left(\Phi_{1}^{\dagger}\Phi_{2}\right)^{2}+\mathrm{h.c.}\right],
\end{aligned}
\end{equation}
where the parameters $\mu_i$ and $\lambda_i$ are real valued and $D_\mu$ is the covariant derivative. We parameterise the 2  doublets as
\begin{equation}\label{phi1-phi2}
\Phi_1=\begin{pmatrix}
G^{+} \\ \frac{1}{\sqrt{2}}(v+h+iG^0)
\end{pmatrix} \qquad \text{and} \qquad
\Phi_2=\begin{pmatrix}
H^{+} \\ \frac{1}{\sqrt{2}}(X+iA),
\end{pmatrix}.
\end{equation}
$G^0$ and $G^\pm$ are the neutral and charged Goldstone bosons, respectively. $v$ is the usual vev of the SM with $v\simeq 246$ GeV. The mass of the $W$ boson is $M_W=g v/2$, $g$ is the $SU(2)$ gauge coupling that relates to the electromagnetic coupling, $e$ as $g=e/s_W$. With $M_Z$ the mass of the $Z$ boson, $M_W^2/M_Z^2=1-s_W^2$ defines $s_W$.

The model has five physical mass eigenstates: together with the CP-even SM Higgs $h$ (with mass $M_h= 125$ GeV) there are two extra neutral scalars, $X$ and $A$, and a pair of charged scalars $H^\pm$.
Depending on the masses of $X$ and $A$, both are possible candidates for DM, nevertheless, the physics is the same through the interchange $(\lambda_5, X) \leftrightarrow (-\lambda_5, A)$. We will consider the scalar $X$ as the lightest extra, neutral, scalar (hence, stable) $M_X < M_A$. With the minimisation condition on the potential, the scalar masses can be used to define $4$ of the parameters of the scalar potential Eqs.~(\ref{VIDM}),~(\ref{phi1-phi2})
\begin{equation}\label{para1}
\begin{aligned}
M_h^2& =3v^3 \lambda_1+\mu_1^2 =-2\mu_1^2 = 2\lambda_1 v^2, \\
M_{H^\pm}^2& = \mu_2^2 + \frac{v^2 \lambda_3}{2},\\
M_X^2& = \mu_2^2 + \frac{v^2 \lambda_L}{2}=M_{H^\pm}^2 +(\lambda_3+\lambda_5)\frac{v^2}{2}, \\
M_A^2& = \mu_2^2 + \frac{v^2 \lambda_A}{2}=M_{H^\pm}^2 +(\lambda_4-\lambda_5)\frac{v^2}{2}=M_X^2-\lambda_5 v^2.
\end{aligned}
\end{equation}
The couplings of the SM Higgs to the new scalars can be used as input physical parameters to re-express one of the independent combinations of parameters of the potential
\begin{equation}\label{paralal}
\lambda_{L/A} =\lambda_3 +\lambda_4 \pm \lambda_5.
\end{equation}
For example, the coupling  between the SM Higgs $h$ and  a pair of lightest neutral scalar $X$ is
\begin{equation}\label{hXX-tree}
 \mathcal{A}_{hXX} = -\lambda_L v \quad (\text{ the quartic coupling}\; hhXX \propto \lambda_L).
\end{equation}
Similarly $h(h)AA \propto \lambda_A , h(h) H^+ H^- \propto \lambda_3$. Observe that $\mu_2$ will be a redundant parameter when $\lambda_L$ is chosen as an input parameter, as we will do. 

There remains an independent parameter to be defined, $\lambda_2$. $\lambda_2$ which  describes the four-point self-interaction within $\Phi_2$ makes this coupling totally elusive when we only consider (at tree-level) interaction of any SM particle. For $W^\pm/Z h$ production at the LHC we are totally insensitive to this parameter, $\lambda_2$. Note that considerations about the stability of the inert vacuum can impose some lower limit on $\lambda_2$. This has first been worked out in~\cite{Ginzburg:2010wa},  reviewed in \cite{Belyaev:2016lok} and mentioned by many authors since, see for example~\cite{Swiezewska:2012ej}. All but one of the benchmark points we will consider have this constraint implemented with the default value $\lambda_2=1$, see later.
                                                                                                                                                                                                                                                                                                                                                 
In turn, we can reconstruct the parameters of the scalar potential though the physical parameters
\begin{equation}\label{para2}
\begin{aligned}
\lambda_1&=\frac{M_h^2}{2v^2}, \\
\lambda_5&=\frac{M_X^2-M_A^2}{v^2}, \\
\lambda_4&=\lambda_5+ 2\frac{M_A^2-M_{H^\pm}^2}{v^2}, \\
\lambda_3&=\lambda_L-\lambda_4-\lambda_5, \\
\mu_2^2& = M_X^2-\lambda_L\frac{v^2}{2} \to \lambda_L=\frac{2(M_X^2-\mu_2^2)}{v^2}.
\end{aligned}
\end{equation}
To sum up,  besides the input parameters of the SM, we will take the set
\begin{equation}\label{inde-paras}
\{M_X,\ M_A,\ M_{H^\pm},\ \lambda_L, \ (\lambda_2) \},
\end{equation}
as the set of  independent input (physical) parameters of the IDM. 

Observe that in the limit of mass degeneracy between the three additional scalars, $M_S=M_X=M_A=M_{H^\pm}$, we have $\lambda_4=\lambda_5=0$ and therefore the Higgs boson has the same coupling to $X,A$ and $H^\pm$ with $\lambda_L=\lambda_3=\lambda_A$. When only $M_A=M_{H^\pm}$, $\lambda_4=\lambda_5$ and hence $\lambda_A=\lambda_3$ but $\lambda_L=\lambda_3+2\lambda_4$. In this case, we write $\Delta M=M_A-M_X>0$. With masses in units of GeV $\lambda_A\sim \lambda_L + \frac{1}{3} (\Delta M/100) ((\Delta M/100 )+2 (M_X/100)) \sim \lambda_L+1$ for $M_X=\Delta M=100$ GeV. $\lambda_A$ increases with increasing $M_X$ and increasing $\Delta M$.
These remarks are useful when we try to understand the deviations from the SM in loop observables. The exchange of heavy scalars would decouple, but heavy scalars can make couplings large. 

It is important to keep in mind that, at  tree-level, the Higgs-strahlung processes ($pp \to W^\pm h, Zh$) we are about to study do not depend on the set of parameters introduced by the new scalars and therefore these processes give exactly the same value for these cross sections at tree-level as in the SM. The NLO QCD corrections are also the same as in the SM, differences appear in the electroweak NLO corrections.

\subsection{The current constraints}
\label{sec::currentconstraints}
The IDM parameter space, Eq.~(\ref{inde-paras}),  is already quite constrained. We have already performed a detailed and thorough investigation of these constraints in \cite{Banerjee:2021oxc}, which we again urge the reader to consult for details. Here we will update the constraints taking into account recent data from key observables. We first redelimit the constraints without assuming that the IDM provides a model of DM. In a second stage, we take the IDM to provide a good DM candidate and derive the more restrictive allowed parameter space. 

\subsubsection{Theoretical constraints}
These concern theoretical arguments based on perturbativity, vacuum stability, unitarity, and charge-breaking minima. We bring no change to the analysis we made in \cite{Banerjee:2021oxc}, see also \cite{Lee:1977eg,Ginzburg:2010wa,Branco:2011iw,Belyaev:2016lok,Swiezewska:2012ej}. In this respect, unless otherwise stated, our default value for $\lambda_2$ is $\lambda_2=1$ as previously stated.

\subsubsection{Experimental constraints}
\label{subsub::exp_constraints}
~
\textbf{1): Electroweak Precision Observables (EWPO)}\\
Through the $S,\ T,\ U $ parameters \cite{Peskin:1991sw}, these relate to the two-point functions of the vector bosons as inferred from precision measurements of $Z$ (and $W$) observables. 
Some of these two-point functions, $\Pi_{AB}(Q^2)$ (for a $A \to B$ transition, $Q^2$ is the momentum transfer of the transition),  $Q^2 \neq M_W^2,M_Z^2$ will appear in the one-loop corrections to $pp  \to Vh$. 
In the IDM, like in most models of New Physics\cite{Peskin:1991sw} and certainly the renormalisable ones, we have verified that $U \simeq 0$
\footnote{3 of our benchmark points have $M_X<M_W$ see Table~\ref{tab::BPs1}-Table~\ref{tab::BPs2}.
The $STU$ description is in principle not valid for such scales. We have verified that the description of the EWPO constraint in terms of only the 3 parameters $S, T, U$ with $U \simeq 0$ is nevertheless still valid in these cases; 
in particular, we have verified that even for the low-scale masses these benchmark points exhibit, the assumption of $U$ being negligible compared to $T$ and/or $S$  still holds. 
We have also verified that the $Q^2$-independent expression for $\Delta S$ ( Eq.~(\ref{eq:deltaST})) reproduces very well the full expression based on the self-energies 2-point functions\cite{Peskin:1991sw} evaluated at $Q^2=M_Z^2$.}.

We can calculate the contribution of the IDM as \cite{Barbieri:2006dq}
\begin{equation}
\label{eq:deltaST}
\begin{aligned}
\Delta T&=\frac{F(M_{H^+}^2, M_A^2) + F(M_{H^+}^2, M_X^2)-F(M_{A}^2, M_X^2)}{32\pi\alpha v^2}, \\
F(x,y)=&
\frac{x+y}{2}-\frac{xy}{x-y}\log\frac{x}{y},  \quad F(x,y)=\frac{(x-y)^2}{6 x},\; \text{for}\; x \sim y, \\
\Delta S&=\frac{1}{2\pi} \int_0^1 x(1-x)\ \log \left[\frac{xM_X^2+(1-x)M_A^2}{M_{H^+}^2}\right]dx. \\
\end{aligned}
\end{equation}

In the fully degenerate limit ($M_X=M_A=M_{H^\pm}$), $\Delta T=\Delta S=0$. In the limit $M_X \ll M_A, M_{H^\pm}$ we have 
\begin{equation}\label{EWPO-dT}
\begin{aligned} 
\Delta T\simeq \frac{1}{24\pi^2 \alpha v^2} M_A \overbrace{(M_{H^\pm} -M_A)}^{\Delta M}\sim 0.05\frac{M_A}{500 \text{GeV}}\frac{\triangle M}{10 \text{GeV}}, \qquad \triangle S \simeq -\frac{5}{72\pi}.
\end{aligned}
\end{equation}
The custodial $SU(2) $ symmetry-breaking parameter, $T$, restricts mass splitting drastically so that one must have $M_A\sim M_{H^\pm}$. 
We take the latest constraints (December 2023) as reported by the PDG~\cite{PDG111,ParticleDataGroup:2024cfk}.

These are more restrictive when setting $U=0$ as fully justified in the IDM. We take the $2 \sigma $ limit with 
\begin{equation}
\begin{aligned}
S&= -0.01 \pm 0.07, \qquad T=0.04 \pm 0.06, \quad \text{with a (strong) correlation } \rho_{ST}=+\ 0.92.\\
\end{aligned}
\end{equation}

\textbf{2): Direct searches at LEP}\\
The direct search constraints on the IDM are obtained by reinterpreting the existing limits in the search for charginos and neutralinos at LEP.
When the masses of the inert scalars are greater than the threshold for LEPII production, these constraints are weaker and can be easily bypassed.  Based on Refs.~\cite{Lundstrom:2008ai,Arhrib:2012ia,Swiezewska:2012eh,Belanger:2015kga}, they can be summarised as follows \cite{Abouabid:2020eik} 
\begin{eqnarray}\label{LEP}
M_{H^\pm} &>&80\ \text{GeV}\ (\text{adopted from the charginos search at LEP-II}), \nonumber\\
\max(M_{A},M_{X}) &>&110\ \text{GeV}\ (\text{adopted from the neutralinos search at LEP-II}),  \\
M_A+M_X &>&M_Z\ \text{from the $Z$ width and} \ M_A+M_{H^\pm} >M_W\ \text{from the $W$ width}.\nonumber
\end{eqnarray}

\textbf{\label{searchesLHC}3): Direct searches at the LHC}\\
The main motivation of this paper is to first present the ingredients of a new tool for performing NLO EW and QCD calculations and to first exploit it to investigate the indirect effects (one-loop) of New Physics in $pp \to Vh$ taking as an example the IDM. We could have limited ourselves to a couple of scenarios (points in the parameter space) of the IDM. Instead, we consider two very large data sets and 6 ``benchmark points". The data sets consist of large scans with A) $ 100\; \text{GeV} < M_X=M_A=M_{H^{\pm}} < 1000\;\text{GeV}$ and B) $200\; \text{GeV} <  M_A=M_{H^{\pm}}=M_X+100\; \text{GeV}< 1000\; \text{GeV}$   (imposing  $0<\lambda_L<8$). The characteristics of the (additional) 6 ``benchmark" points are given in Table~\ref{tab::BPs1} and Table~\ref{tab::BPs2}. The details of the impact of the direct searches at the LHC on the parameter space we have chosen to illustrate our calculations are exposed in our rather long Appendix. This is to avoid disrupting the core flow of the paper. First of all, the large scan A) is the all-degenerate scenario which passes the LHC constraint trivially: no decay of any IDM state to another is possible. 
One can not trigger on any decay product; this will remain the case for any future LHC data. As for the other model points, to summarise,  we have found that the constraints, on our sets of parameter space, can be beautifully arrived at on the basis of a reinterpretation using the very important and critical electroweakino production\cite{CMS:2017moi,validation:electroweakinos} data. The overwhelming importance of this analysis is fully justified in the Appendix and in \cite{Banerjee:2021oxc}. Moreover, three of the benchmark points (BP0, BP3, BP4) are subject to a full recast that takes advantage of a host of LHC processes / analyses. The recast is an extension and update (with the most recent LHC data) of a previous detailed study\cite{Banerjee:2021oxc} performed using (validated) {\tt MadAnalysis 5}~\cite{Conte:2012fm,Conte:2014zja,Dumont:2014tja,Conte:2018vmg,Araz:2020lnp} analyses. This updated recast confirms that the results that would have been based solely on the reinterpretation using electroweakino production\cite{CMS:2017moi,validation:electroweakinos} are robust and serve also as further validation of the reinterpretation analysis. We do not expect that including the full Run 2 data and even later data will have an additional impact. One of the main reasons is the fact that our points that were not subjected to a complete recast (with full LHC data)  feature small $\Delta M$ (scan B for example) and have too small cross sections. Allowing (much) larger $\Delta M$ than considered in our study might call for a fuller recast; however, EWPO do not allow such scenarios. Besides, again, the purpose of our paper and its novelty is to point to the importance of investigating  the (loop)  indirect effects of New Physics in LHC processes.

\begin{figure}[!ht]
\centering
\includegraphics[scale=1.3]{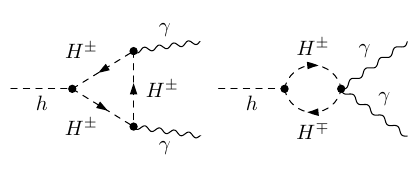}
\caption{The new Feynman diagrams introduced by the IDM for  $h \to \gamma \gamma$ process at the one loop  level.}
\label{fig::hgamgamvert}
\end{figure}

\textbf{4): The LHC Higgs to diphoton, $\mu_{\gamma \gamma}$}\\
The diphoton decay of the SM Higgs boson ($h \to \gamma \gamma$), see Fig.~\ref{fig::hgamgamvert}, is also an important restriction. It receives a contribution from the charged scalars and therefore restricts the $hH^+H^-$ coupling, $\lambda_3$ and the mass, $M_{H^\pm}$. Similar contributions appear in $pp \to Vh$ and can therefore be severely constrained.

The ratio of diphoton signal strength in the IDM is defined as
\begin{equation}
\begin{aligned}
\mu_{\gamma \gamma}&=\frac{\text{Br}_{IDM}(h \to \gamma \gamma)}{\text{Br}_{SM}(h \to \gamma \gamma)}.
\end{aligned}
\end{equation}
The theoretical expression for $\text{Br}_{IDM}(h \to \gamma \gamma)$ can be found in Refs.~\cite{Djouadi:2005gi,Arhrib:2012ia} and the branching fraction in the SM is taken as $\text{Br}_{SM}(h \to \gamma \gamma)=(2.27\pm 0.05)\times 10^{-3}$.
The Higgs diphoton signal strength from ATLAS and CMS has been reported with $\mu_{\gamma \gamma}^{\text{ATLAS}}=1.04_{-0.09}^{+0.10}$~\cite{ATLAS:2022tnm} and $\mu_{\gamma \gamma}^{\text{CMS}}=1.12 \pm 0.09$~\cite{CMS:2021kom}. 
We impose the combined limit
\footnote{
For the combination of ATLAS and CMS on $\mu_{\gamma \gamma}$  we use the weighted average  formula as recommended in the PDG (see Eq. 40.8 in Ref.~\cite{PDG222}
We use 
\begin{eqnarray}
\frac{1}{\bar{\sigma}^2}=\sum_i \frac{1}{\sigma_i^2}, \quad \text{and}  \quad \frac{\bar{\mu}}{\bar{\sigma}^2}= \sum_i \frac{\mu_i}{\sigma_i^2}
\end{eqnarray}
where $\mu_i$ and $\sigma_i$ are the individual signal strengths and their 1-$\sigma$ uncertainties respectively.
We approximate the asymmetric error of ATLAS with a symmetric one taking $\sigma_1=\sigma_{\text{ATLAS}}=(0.1+0.09)/2=0.095$ ($\sigma_2=\sigma_{\text{CMS}}=0.09$).
}
\begin{equation}\label{Rgg-exp}
\begin{aligned}
\mu_{\gamma \gamma} =1.08\pm 0.065.
\end{aligned}
\end{equation}

\textbf{5): Invisible Higgs decay}\\
This is of relevance only for $M_X < M_h/2 \sim 62.5$ GeV. 
The invisible branching fraction for $h \to XX$ process at the tree-level is expressed as
\begin{equation}\label{IN::BR}
\begin{aligned}
\text{Br}(h \to XX)&= \frac{1}{\Gamma_h^{\text{SM}}} \frac{\lambda_L^2 v^2}{32\pi M_h} \sqrt{1-\frac{4M_X^2}{M_h^2}} \\
&=\left[1.183(\lambda_L\times 10^{3})^2 \sqrt{1-0.64(M_X/50)^2} \right] \times 10^{-3} \\
&=2.83\times 10^{-3}\quad \text{for} \quad M_X=57 \text{ GeV}, \quad \lambda_L=2.4\times 10^{-3},
\end{aligned}
\end{equation}
where the total width of SM Higgs is taken $\Gamma_h^{\text{SM}}=4.07 \pm 0.16$ MeV. The ATLAS and CMS constraints are based on a combination of searches~\cite{ATLAS:2023tkt,
CMS:2023sdw}. We incorporate the limit set by ATLAS \cite{ATLAS:2023tkt} 
\begin{equation}\label{atlas-invi-higgs}
\begin{aligned}
\text{Br}^h_{{\rm inv.}}< 0.107.
\end{aligned}
\end{equation}

\textbf{6): Models A with no DM constraints, benchmark points}

While we will be studying deviations from the SM in the $Vh$ cross sections, across a large allowed parameter space taking all the constraints we have just listed (not assuming that $X$ is a DM candidate), we will take three representative benchmark points (BP's) to analyse kinematical distributions.  These 3 benchmark points are listed in Table~\ref{tab::BPs1}. BP1 features a fully degenerate spectrum, all extra scalars are relatively heavy but share the same mass, $\lambda_L$ is also relatively large. The masses in BP2 are not as heavy, $A$ and $H^+$ are degenerate in mass, and there is a $100$ GeV mass splitting between $X$ and $A/H^+$. BP0 has $M_X$ light and the other two scalars are degenerate and quite heavy; however $\lambda_L$ is smaller than in the other two benchmark points. 
\begin{table}[!htb]
\renewcommand\arraystretch{1.1}
\begin{center}
\begin{tabular}{| p{2cm}<{\centering} |p{2cm}<{\centering} |p{2cm}<{\centering} |p{2cm}<{\centering} |p{2cm}<{\centering}|p{2.2cm}<{\centering} |}
\cline{2-6}
\multicolumn{1}{c|}{}& $M_{X}$(GeV)   & $M_{A}$(GeV)   & $M_{H^\pm}$(GeV)   & $\lambda_L$  & ($\lambda_3,\lambda_A$)  \\
\hline 
BP0  & 70             &   571       &          571      & 1.0    &(11.61,11.61)            \\
\hline
BP1   & 500             & 500           & 500                & 7.08    &(7.08,7.08)           \\
\hline
BP2   & 220             & 320           & 320                & 2.31  & (4.09,4.09)           \\
\hline
\end{tabular}
\end{center}
\caption{In case the IDM is not a model for DM, three benchmark points selected to display differential distributions for $Vh$ production at the LHC. For BP0 we take $\lambda_2=2$.
As pointed out in the text, the value of $\lambda_2$ has no impact on the electroweak NLO calculation.
}
\label{tab::BPs1}
\end{table}

\textbf{7): Dark Matter constraints}

\textbf{a): Dark matter Relic Density.}
The lightest neutral scalar $X$ could be a good candidate for DM. To qualify as such, it must meet the requirement of the relic density, $\Omega h^2$, which we calculate within the freeze-out scenario. We calculate such contributions with {\tt micrOMEGAs}~\cite{Belanger:2001fz,Belanger:2004yn,Belanger:2006is,Belanger:2013oya,Belanger:2018ccd}.  {\tt micrOMEGAs} takes the annihilation cross sections at tree-level, albeit with an effective (electromagnetic) coupling at $\alpha(M_Z)\sim 1/128.07$ instead of the on-shell coupling in the Thomson limit $\alpha \sim 1/137.036$. However, we have shown in~\cite{Banerjee:2019luv,Banerjee:2021oxc,Banerjee:2021anv,Banerjee:2021xdp,Banerjee:2021hal} that radiative corrections are not negligible. We therefore  include a theoretical uncertainty of 20\% to be added, linearly, to the experimental  uncertainty~\cite{Planck:2018vyg} on the relic density determination and require
\begin{equation}
\label{eq:relicdensitybound}
\begin{aligned}
0.095< \Omega h^2 < 0.143
\end{aligned}
\end{equation}

\textbf{b): Dark matter direct detection}
The spin-independent DM nucleon cross section proceeds through the SM Higgs. It can be written as 
\begin{equation}\label{Con::DMdirect}
\begin{aligned}
\sigma_{SI}^h \sim 
\left(\frac{\lambda_L 10^3}{M_X/100\text{GeV}} \right)^2\times 8.53\times 10^{-49}\ \text{cm}^2.
\end{aligned}
\end{equation}
This gives a strong constraint on $\lambda_L/M_X$. We have updated this constraint using the latest data given by the LZ experiment in 2022 \cite{LZ:2022ufs}. The constraint can be approximately cast as 
\begin{equation}\label{Con::Xenon1T}
\begin{aligned}
\sigma_{SI}^{\text{LZ}} < 3 \times 10^{-47} \cdot  \left(M_X /100\text{GeV} \right)\ \text{cm}^2.
\end{aligned}
\end{equation}
for a DM mass in the region $40 < M_X <100$ GeV, slightly improving until the minimum at $30$GeV. We have imposed the constraint 
\begin{equation}\label{DD::Lmax}
\begin{aligned}
|\lambda_L|\times 10^3 <6 \times (M_X /100\ \text{GeV})^{3/2}\quad \to\quad |\lambda_L|<|\lambda_L^{\text{max}}|=0.0035,\ \text{for}\ M_X=70\ \text{GeV}.
\end{aligned}
\end{equation}

Some IDM scenarios may lead to a lower relic density than what is set by the bound in Eq.~(\ref{eq:relicdensitybound}). These would be acceptable but only furnish a fraction of the DM in the Universe, and the IDM would be part of a larger picture that provides the rest of the observed DM relic density. 
In this case, the direct detection rate due to the IDM DM is smaller due to the smaller IDM DM halo fraction. Indeed, the detection rate is directly proportional to
$\rho^{{\text DM}}_{\odot} \sim 0.3 \text{GeV}/\text{cm}^3  $, the local DM density to which the IDM DM contributes only a fraction. 
In this case, we assume that this fraction is the same as the one on cosmological scales which is set by the relic density (see, for example, \cite{Boudjema:2012in}). In this case, we rescale the density in Eq.~(\ref{Con::DMdirect}) to arrive at 

\begin{equation}\label{con-invi}
\begin{aligned}
\lambda_L < \lambda_L^{max} \sqrt{\frac{\Omega_{DM}^{\text{Planck}} h^2}{\Omega_X h^2}},\end{aligned}
\end{equation}
where  $\Omega_X h^2 \equiv \Omega_X^{{\text IDM}} h^2 $ and $\Omega_{DM}^{\text{Planck}}h^2=0.1191$  denote the relic density of the IDM and the one extracted from the Planck data, respectively. Because the former is greater than the latter,  it is clear that the restriction from Eq.~(\ref{con-invi}) is weaker than Eq.~(\ref{IN::BR}). We do nevertheless apply this constraint only if the IDM provides no less that $50\%$ of the observed DM. Otherwise the model is considered as not having a connection to DM. 

In the case where $M_X \sim M_A \sim M_{H^\pm} \gg M_W$ there are large one-loop electroweak corrections to the scattering cross section \cite{Cirelli:2005uq,Klasen:2013btp} that can be dominant\footnote{In \cite{Abe:2015rja} beyond tree level calculations were performed in the rather constrained case with $M_h \simeq 2 M_X$.}. The latter cross section only depends on the gauge coupling and the Higgs mass. In these limiting cases, the cross section is driven by the gauge interaction and depends only on the SM parameters. With $M_h=125$ GeV the one-loop result gives \cite{Banerjee:2019luv,Cirelli:2005uq}
\begin{equation}\label{DD-gauge-heavy}
\begin{aligned}
\sigma^{g}_{Xn}=8.88 \times 10^{-47}  \text{cm}^2.
\end{aligned}
\end{equation}
We take this to be a valid approximation for $M_X > 300$ GeV. In this range of masses this contribution passes the LZ2022 limit. For the tree-level contribution to be larger one needs 
\begin{equation}
    \begin{aligned}
        \lambda_L \times 10^2 >  M_X \text{(GeV)}/100.
    \end{aligned}
\end{equation}

We thoroughly studied the allowed parameter space when DM constraints are imposed beyond tree-level considerations. We again urge the reader to refer to \cite{Banerjee:2021oxc} for more details. After all constraints are updated, two windows survive: 
\begin{itemize}
\item A low mass, $M_X< M_W$,  DM candidate \cite{Banerjee:2021oxc} where various annihilation/co-annihilation processes were studied beyond tree-level \cite{Banerjee:2021anv, Banerjee:2021hal, Banerjee:2021xdp}. 
\item A high mass $M_X> 500$GeV region. An analysis beyond tree-level in this region is performed in \cite{Banerjee:2019luv}. 
\end{itemize}
There is no intermediate mass range $M_W< M_X<500$ GeV DM candidate in the IDM. The key characteristic for all possible DM candidates in both allowed mass ranges is that $\lambda_L$ is at least two orders of magnitude smaller than what we considered for the IDM without the DM constraints
\footnote{
We thank the referee for pointing to us that our results for the allowed mass regions of the IDM agree with those in \cite{Kalinowski:2020rmb}. In the latter, the theoretical errors on the calculation of the relic density were neither calculated nor estimated. 
}.

\textbf{8): Models B with added DM constraints, benchmark points}
\begin{table}[!htb]
\renewcommand\arraystretch{1.1}
\begin{center}
\begin{tabular}{| p{2cm}<{\centering} |p{2cm}<{\centering} |p{2cm}<{\centering} |p{2cm}<{\centering} |p{2cm}<{\centering}|p{2cm}<{\centering} |}
\cline{2-6}
\multicolumn{1}{c|}{}&$M_{X}$(GeV)   & $M_{A}$(GeV)   & $M_{H^\pm}$(GeV)   & $\lambda_L(10^{-3})$ & ($\lambda_3,\lambda_A$)  \\
\hline
BP3  & 59              & 113           & 123                & 1.0      &(0.38, 031)          \\
\hline
BP4   & 60              & 68            & 150                & 0.0     & (0.62,0.03)           \\
\hline
BP5  & 550             & 551           & 552                & 19.3    &(0.09,0.06)         \\
\hline
\end{tabular}
\end{center}
\caption{Three benchmark points that satisfy all constraints including the DM constraint.}
\label{tab::BPs2}
\end{table}

While we will scan over a large parameter space, see paragraph {\it Direct searches at the LHC} in Section~2), Table~\ref{tab::BPs2} 
lists (equivalent of Table~\ref{tab::BPs1}) our three benchmark points that are good DM candidates. These are taken from our studies in \cite{Banerjee:2021anv} and \cite{Banerjee:2016vrp}.


\section{Higgs-strahlung  in the SM: QCD and EW 1-loop corrections}
\label{sec::vhsm}
\subsection{A recap of precision calculations in the SM for $Vh$ processes at the LHC}
Precision calculations of Higgs-strahlung processes in the SM span a period of almost four decades. 
These calculations cover both QCD calculations
\cite{Barger:1986jt,Dicus:1988yh,Kniehl:1990iva,Han:1991ia,Ohnemus:1992bd,Baer:1992vx,Zecher:1994kb,Cao:1998br,Harlander:2002wh,Brein:2003wg,Brein:2011vx,Ferrera:2011bk,Banfi:2012jh,Brein:2012ne,Altenkamp:2012sx,Englert:2013vua,Ferrera:2013yga, Kumar:2014uwa,Ferrera:2014lca,Hespel:2015zea,Campbell:2016jau,Hasselhuhn:2016rqt,Caola:2017xuq,Ferrera:2017zex,Harlander_2018,Gauld:2019yng,Majer:2020kdg,Heinrich:2020ybq,Chen:2020gae,Wang:2021rxu,Baglio:2022wzu,Degrassi:2022mro,Chen:2022rua} and electroweak corrections \cite{Ciccolini:2003jy,Denner:2011id,Denner:2011rn,Denner:2014cla,Granata:2017iod,Obul:2018psx}. 
The improvements in the calculations within the SM extended the results of inclusive cross sections to exclusive distributions including $W/Z$ and Higgs decays. 
The state of the art is the relatively recent computation of the N$^3$LO \cite{Baglio:2022wzu} and the higher order correction to the $gg$ induced $ZH$ channel \cite{Davies_2021,Chen_2021,Alasfar_2021,Wang_2022,Chen_2022,Degrassi:2022mro,Chen:2022rua}. Although the latter is a one-loop process in leading order, it nevertheless represents about $6\%$ of the yield due to the importance of the $gg$ luminosity at the LHC.  
At LO, it suffers from a large factorisation scale uncertainty, hence the need for the new calculations. 
As is known, the largest corrections for $Vh$ are the NLO QCD corrections which are the same as those in Drell-Yan NLO production, higher-order (QCD) corrections are comparatively quite modest but
reduce quite considerably the scale variation.
In contrast, the NLO EW corrections are in the order of a few per cent but nonetheless quite important, making these processes ideal places to study variations due to New Physics.

\subsection{Our general approach to NLO QCD and electroweak corrections }

Although our motivation is to weigh how much the contributions from the IDM in $Vh$ production at LHC can be, it is important to show that our code can perform automatic calculations both in the SM and in the IDM (as prototype of a New Physics model). 
The implementation of the different contributions of the radiative corrections in the code borrows from textbook techniques that are put together to serve as checks of the result at different steps of the automation process. 
Such a combined study allows us to show how different contributions and subprocesses ($qq, \gamma q$-induced,..) are impacted as well as to give a measure of the scale and PDF uncertainties. 

\subsection{Ingredients of our code, {\tt SloopS}}
Our calculation of the different contributions to $Vh$ production at the LHC is based on the automatic tool {\tt SloopS}~\cite{Boudjema:2005hb,Baro:2007em,Baro:2008bg,Baro:2009na,Boudjema:2011ig,Boudjema:2014gza,Belanger:2016tqb,Belanger:2017rgu,Banerjee:2019luv} for the computation of cross sections (and distributions) and decay widths. {\tt SloopS} relies on {\tt LanHEP} \cite{Semenov:2014rea,Semenov:2010qt,Semenov:2008jy,Semenov:2002jw,Semenov:1998eb,Semenov:1996es} for the definition of the model. It relies on the bundle of packages based on {\tt FeynArts}\cite{Kublbeck:1990xc}, {\tt
FormCalc}\cite{Mertig:1990an} and {\tt LoopTools}\cite{Hahn:1998yk} for one-loop calculations. The one-loop library has, since \cite{Baro:2007em} been continually augmented with in-house routines and improvements. {\tt LanHEP} implements the complete set of all (ultraviolet) counterterms of the model so that full one-loop renormalisation is implemented. For the electroweak sector this can be the on-shell (OS) or a mixed \msbar-OS  scheme. We will come back to these issues. Our implementation of the IDM model (including its  (UV) renormalisation)\footnote{Users can also access the working files and examples as explained on the wikipage below\\
$~\quad\ \ $ {\tt https://lapth.cnrs.fr/projects/PrecisionCalculations/idm\_at\_1loop}
} has been detailed in \cite{Banerjee:2021oxc}. The code permits a first internal check, by testing the gauge parameter independence of the result of any cross section\cite{Banerjee:2021oxc}.
For QCD, with the processes at hand, a full  standard \msbar  scheme is taken with a renormalisation scale $\mu_{\text {R}}$ and a strong coupling constant defined as $\alpha_s(\mu_{\text {R}}^2)$. The latter is calculated according to \cite{Chetyrkin_2000} with $n_f=5$ active flavours and 4-loop beta function \cite{Buckley:2014ana}\footnote{A 2-loop $\beta$ function would have been sufficient in accordance with the NLO PDF evolution.}. 

\subsection{Soft and Collinear singularities }
Until now \sloops has been exploited for EW corrections to the SM and a host of renormalisable BSM extensions and also QCD corrections for DM calculations. The latter were essentially QED type corrections for which the infrared soft singularities are captured with a vanishingly small photon mass and collinear singularities further regulated with fermion masses. We have now extended \sloops so that applications to hadronic collisions are possible. In particular, one can now exploit \sloops to handle {\it massless partons}. 
First, for the renormalisation of the electroweak sector, we use   $G_\mu$   as an input  parameter {\it in lieu} of the electromagnetic fine structure constant defined in the Thomson limit.  The other important extension is the use of dimensional regularisation to handle infrared and collinear singularities for the virtual and the radiation (real) parts, examples of which will be detailed for each subprocess contributing to $Vh$ production. Dimensional regularisation with
$ d=4-2\epsilon_{\text{IR}}$ is applied to both the QCD and electroweak (QED) infrared/collinear  singularities. To have extra (redundant) checks on this part of the calculation, we deploy two implementations. and make sure that they return the same result. Work on automating the treatment of real radiation for processes more complex than, say, $Vh$, is still ongoing. For what concerns us here, we briefly describe how the real corrections are implemented through two methods that will be compared numerically when we study $Vh$ production.

\subsubsection{The two-cutoff phase space slicing}
\label{sec::tcpss}
For the $qq$-initiated subprocess for $Vh$ production, the virtual part, $\hat{\sigma}_V$,  consists of a $2 \to n$, process with $n=2$, and the radiation part, $\hat{\sigma}_R$, with the emission of an extra parton being a $2 \to n+1=3$ process. We closely follow the two-cutoff phase space slicing method, TCPSS\cite{Harris:2001sx}. It introduces, generally, two parameters to control the infrared singularities ($\delta_s$) and collinear($\delta_c$) singularities. The phase space is decomposed into a soft infrared region, S, where the energy of the extra parton, here a gluon $g$, is required to satisfy $ E_g < \delta_s \sqrt{\hat{s}}/2$. $\hat{s}$ is the partonic Mandelstam variable. S lends itself to an analytical evaluation that is automatically fed into the code.
{$\delta_s \ll 1$ is chosen as small as possible to allow the general analytical approximation to be valid. The hard region H, $ E_g > \delta_s \sqrt{\hat{s}}/2$, generally contains collinear singularities. This region H is further subdivided into a (hard) collinear region ($HC$)  and a hard non-collinear region ($H\bar{C}$). This subdivision is set by a cut $\delta_c$ on the virtuality of a pair of partons. For our processes, for example, with $i,j$ being the initial partons, the requirement is $ (\hat{s}_{ig},\hat{s}_{jg}) \leq \delta_c \hat{s}, \hat{s}_{ij} = (p_i+p_j)^2$. 
Our default implementation is $\delta_c=\delta_s/50$. 
$H\bar{C}$ is amenable to a Monte Carlo integration of the $2 \to 3$ $qq$-initiated subprocess in $d=4$. 
For the numerical integration of the hard non collinear ($H\bar{C}$) region to be stable, $\delta_s$ cannot be taken infinitely small. Therefore, we usually take different values of small enough $\delta_s$  until the result of adding the soft contribution and the hard contribution no longer depends on the small value of $\delta_s$ and a plateau in $\delta_s$ is reached for the combined contributions. Although the ingredients to build up the method can be automated, the fact that on has to {\it observe} stability in $\delta_s$ makes full automation not so.
We will show explicitly how precisely a $\delta_s$ ($\delta_c)$ independent result is obtained for both the QCD and QED corrections. 
The $2 \to 3$ $ H\bar{C}$ Monte Carlo integration is the most time-consuming part. It requires, comparatively, higher statistics than the other parts that enter the evaluation of the NLO corrections, either QCD or QED. 

To sum up, at the partonic level (hatted $\hat{\sigma}$), the IR/Collinear structure of the NLO cross sections in the TCPSS method can be written as 
\begin{eqnarray}\label{TCPSS}
\hat{\sigma}^{\text {NLO}}&=& \hat{\sigma}^{\text LO}+\Delta\hat{\sigma}^{\text {NLO}}, \quad {\rm}  \\
\Delta \hat{\sigma}^{\text {NLO}} &=&\Delta \hat{\sigma}^{\text {NLO}}_{\text{TCPSS}}=\hat{\sigma}_{\text{V}}(\frac{1}{\epsilon_{\text{IR}}}, \frac{1}{\epsilon^2_{\text{IR}}}) + \overbrace{\hat{\sigma}_{\text{S}}(\delta_s,\frac{1}{\epsilon_{\text{IR}}}, \frac{1}{\epsilon^2_{\text{IR}}})  + \hat{\sigma}_{\text{H}  \text{C}}(\delta_s,\delta_c,\frac{1}{\epsilon_{\text{IR}}})+ \hat{\sigma}_{\text{H}\overline{C}}(\delta_s,\delta_c)}^{\hat \sigma_R},\nonumber
\end{eqnarray}
$\hat{\sigma}^{\text LO}$  is the leading order (differential) cross section at the partonic level. 
The virtual correction $ \hat{\sigma}_V$  includes UV renormalisation.
For $Wh$ production, compared to the QCD corrections where the final state has no colour charge, photon radiation off electrically charged  $W$ 
(final state) gives rise to an (additional) infrared, soft, but non-collinear singularity because of the mass of the $W$, see later Sect.~\ref{sub:EWreal}. The singularity is contained in the QED part of $\hat{\sigma}_V$ and $\hat{\sigma}_S$, where the single-pole ($1/\epsilon_{\text{IR}}$)  dependence cancel each other out as does the remaining $\delta_s$ dependence between $\hat{\sigma}_S$ and $\hat{\sigma}_{\text{H}\overline{C}}(\delta_s,\delta_c)$ where $\delta_c=0$ (there is no collinear singularity). \\

In $Vh$ production the collinear singularity in $\hat{\sigma}_{\text{H}  \text{C}}(\delta_s,\delta_c,\frac{1}{\epsilon_{\text{IR}}})$ concerns radiation from the initial state. As is known, see~\cite{Harris:2001sx} for example, this singularity is absorbed into the bare parton distribution functions (PDFs) leaving a finite remainder, which is written in terms of modified parton distribution functions. This important procedure is standard; see \cite{Harris:2001sx}.
It relies on factorisation in the HC term with the help of the corresponding DGLAP kernels. After convolution with the PDF, the full partonic cross sections are elevated to physical cross sections, $\sigma^{\text{LO}} $ and $  \sigma^{\text{NLO}}$ (with no dependence on $\delta_s, \delta_c$).
There remains, of course, the dependence on the renormalisation scale, $\mu_R$, and the factorisation scale, $\mu_F$.

In addition to the corrections to the subprocess initiated by $qq$, a complete order $({\cal{O}}(\alpha,\alpha_s))$ involves the production induced by $qg,q\gamma$. Although soft infrared finite, these contain a collinear singularity which is treated in a similar manner to the $qq$ case: dimensional regularisation for the HC and redefinition of the PDF with the corresponding splitting functions. The $H\bar{C}$ are calculated numerically in 4-dimensions. We will also check the $\delta_c$ independence of the total result of this contribution, which is written as Eq.~(\ref{TCPSS}) but with $\sigma_V=\sigma_S=0$ and $\delta_s \to 0$ in the last two terms. \\
The other corrections, $gg,\gamma \gamma$-induced production, are finite ( both in the UV and the IR sense) in an NLO approach.\\

\subsubsection{Dipole subtraction}
\label{sec::dipolesub}
The other approach for dealing with infrared/collinear singularities is the more efficient dipole subtraction (DS) method. It is our default implementation. The efficiency and accuracy of the method lies in the fact that subtraction is essentially achieved before integration over the phase space is performed. Moreover, it does not rely on an approximation in which one has to quantify how small a parameter such as $\delta_s$ should be. 
In our code, DS is carried out {\it à} la Catani-Seymour \cite{Catani:1996jh,Catani:2002hc,Catani:1996vz,Dittmaier:1999mb,Nagy:2003tz,Nagy:1998bb,Frederix:2008hu,
Gehrmann:2010ry}. For the EW corrections, we adapt the dipole formulae in QCD provided in Refs. \cite{Catani:1996jh,Catani:2002hc,Catani:1996vz} in a straightforward way (change of gauge coupling/colour factor) to the case of dimensionally regularised photon emission.
\\

As is well known, the basic idea of DS is to construct a simple local counterterm $d\hat{\sigma}_A$ that captures the {\it universal structure} of the infrared/collinear singularity of the radiation. This local term will be subtracted from the real part and added back, appropriately,  to the virtual part. The  structure of the radiation  
is encapsulated into {\it universal} dipoles $d V_{\text{dipole}}$ that enter as factors of the tree-level (LO) cross section. We will have to sum over all possible radiations. Our implementation of the subtraction includes a ``cut" parameter, $\alpha^\prime$ \footnote{This corresponds to the parameter $\alpha$ in the original proposal \cite{Nagy:2003tz,Nagy:1998bb}, introduced to control the size of the dipole phase space. In our code $\alpha$ is reserved for the electromagnetic fine structure constant.}. In contrast to $\delta_{s,c}$, any value of $\alpha^\prime$, $0<\alpha^{\prime} \leq 1$ serves only as an additional check: the $\alpha^\prime$
independence of the final result. The method lends itself to full automation since we do not need to adjust the parameter $\alpha^{\prime}$ to achieve stability of the result. 
For a (LO) $ 2 \to 2 $ process (requiring $2 \to 3$ radiation ) like the case at hand, in the dipole implementation, Eq.~(\ref{TCPSS}) turns into

\begin{equation}\label{dipole}
\begin{aligned} 
\Delta \hat{\sigma}^{\text {NLO}}&=
\Delta \hat{\sigma}^{\text {NLO}}_{\text{Dipole}}= \hat{\sigma}_V + \hat{\sigma}_R = \bigint_2 \left(d \hat{\sigma}_V + \underbrace{ \int_1d \hat{\sigma}_A}_{d \hat{\sigma}_{\overline{\text{IntDipole}}}} \right)+ \bigint_3 \underbrace{ \left(d \hat{\sigma}_R -d \hat{\sigma}_A \right)}_{d \hat{\sigma}_\text{Dipole}}, \\
&=\left[\hat{\sigma}_V(\frac{1}{\epsilon_{\text{IR}}}, \frac{1}{\epsilon^2_{\text{IR}}}) + 
\hat{\sigma}_{\overline{\text{IntDipole}}}(\alpha^\prime,\frac{1}{\epsilon_{\text{IR}}}, \frac{1}{\epsilon^2_{\text{IR}}})\right] \;+\;
\hat{\sigma}_{\text{Dipole}}(\alpha^\prime), 
\quad \text{with}\\
d\hat{\sigma}_A &= \sum_{\text{dipoles}} d \hat{\sigma}^\text{LO} \otimes d V_{\text{dipole}}.
\end{aligned}
\end{equation}
The subscripts for the integrals refer to the dimensionality of the phase space. The $2 \to 3 $ can be performed numerically (with $d=4, \epsilon_\text{IR}\to 0$). $\otimes$ represents phase space convolution and sums over colour and spin indices.  Similarly to the TCPSS method, the collinear initial state singularity is absorbed in the (re)-definition of the PDF and also involves the splitting kernels. The initial-state collinear singularity requires a subtraction term, $\hat{\sigma}_C(1/\epsilon_{\text{IR}},\mu_F^2)$, to be added to the regulated $2 \to 2$ phase space, $d \hat{\sigma}_V+d \hat{\sigma}_{\overline{\text{IntDipole}}}$, Eq.~(\ref{dipole}), see \cite{Catani:1996vz,Gleisberg:2007md} for details,(with $\sigma_C$ included, $d \hat{\sigma}_V+d \hat{\sigma}_{\overline{\text{IntDipole}}} \to d \hat{\sigma}_V+d \hat{\sigma}_\text{IntDipole}$).
We have to take into account all possible partonic cross sections, so $\hat \sigma$ above for a particular subprocess  is  $\hat \sigma_{ij}$, where $i$ is the incoming parton. The LHC $PP$ cross section writes
\begin{equation}
\begin{aligned}
 \sigma^{PP \to Vh}(p_1,p_2) = \sum_{ij} \int {\rm d}x_1 \; {\rm d}x_2 \; & G_i^P(x_1, \mu_F^2) G_j^P(x_2, \mu_F^2)
 \bigg( \\
 &d\hat{\sigma}_{ij}^{\rm LO} (x_1 p_1, x_2 p_2)+d\hat{\sigma}_{ij}^{\rm NLO}(x_1 p_1, x_2 p_2, \mu_F^2) \bigg) 
\end{aligned}
\end{equation}
$P$ is the proton, $p_{1,2}$ are the momenta of the incoming  protons at the LHC. $i,j$ are the initial partons. $x_{1,2}$ are the Bjoerken $x$,  $G_i^P(x_1, \mu_F^2)$ is the scale-dependent PDF of parton $i$ inside the proton, and $\mu_F$ is the factorisation scale. Note that the NLO cross section has a $\mu_F^2$ dependence brought about by counterterm/subtraction of the collinear singularity that redefines the PDF. The integral is over the PDFs and phase space. 

For all three $pp \to W^\pm h, Zh$ processes,  the NLO QCD correction to $q \bar q^{(\prime)}$ only involves an initial (emitter)-initial (spectator) type contribution. When generalising to the EW (QED) corrections, in the $W^\pm h$ case,  extra combinations of dipoles are needed to deal with the infrared singularity from photon radiation off the charged (massive) $W^\pm$.

As we will demonstrate, the handling of the infra-red/collinear singularities for $pp \to Vh$ performs extremely well. Not only does $\alpha^\prime$-independence serve as an internal test of DS, as is the {\it plateauing} in $\delta_s$ and $\delta_c$ for TCPSS, but the agreement between the two methods is an extra independent test.

\subsection{Standard Model Parameters}
This is a good place to define and set the SM parameters that are relevant for the processes that we will study and quantify.

The Standard Model input parameters will be taken as \cite{ParticleDataGroup:2022pth}
\begin{equation}\label{paras-SM}
\begin{aligned}
M_Z &=& 91.1876\ \text{GeV}, \qquad M_W = 80.385\ \text{GeV}, \qquad M_h=125\ \text{GeV} \\
 M_t &=& 173.21\ \text{GeV}, \qquad G_\mu = 1.1663787 \times 10^{-5}\  \text{GeV}^{-2}. \qquad \qquad
\end{aligned}
\end{equation}
All fermions, but the top, are taken  massless. 
The Cabibbo-Kobayashi-Maskawa (CKM) matrix is taken to be the unit matrix. We will use different
PDF sets with $\alpha_s(M_Z)= 0.118$ via LHAPDF6 \cite{Buckley:2014ana} in the $n_f = 5 $ fixed-flavour scheme in both the LO and the NLO predictions. 
We will take the renormalisation, $\mu_R$, and factorisation, $\mu_F$, scales equal and track, as is done in Ref.~\cite{Baglio:2022wzu} the scale dependence through the parameter $x_{R/F}$ defined as 
\begin{equation}\label{renox}
\begin{aligned}
\mu_R=\mu_F=x_{R/F}\;(M_V + M_h).  
\end{aligned}
\end{equation}
As in Ref.~\cite{Baglio:2022wzu} we will vary $x_{R/F}$ in the range $1/8\leq x_{R/F}\leq5$. 
This is a much wider range than is generally considered for NLO corrections; it translates into scales ranging from $~25$ GeV to almost 1 TeV.
Unless otherwise stated, the cross sections refer to the use of the {\tt NNPDF31-luxQED} PDF set \cite{NNPDF:2017mvq,Bertone:2017bme}. 
We will then take $x_{R/F}=2$ for reasons we will make explicit in due course~\cite{Baglio:2022wzu}. 
All our figures are for a run performed for the LHC at $\sqrt{s}=13$ TeV.

\subsection{Leading Order, LO}
In the five (massless) flavour scheme, associated Higgs-vector boson production, Higgs-strahlung, at LO (partonic) is of a Drell-Yan type
\begin{equation*}
\begin{aligned}
q_{i}\ \bar{q}_{j} \to V h, \quad V=Z, W^\pm,  
\end{aligned}
\end{equation*}
where  the $q_i, q_j= u,d,s,c,b$. It is shown in Fig.~\ref{fig::tree}.

\begin{figure}[!hbt]
\centering
\includegraphics[scale=0.9]{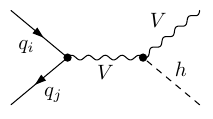}
\caption{Tree level Feynman diagrams for Higgs-strahlung process in the five flavour scheme. The $q_{i(j)}$ represents all (non top) quark flavours and anti-quarks and $V=Z,\ W^\pm$.}
\label{fig::tree}
\end{figure}
Observe again that, at this first order, the cross section is exactly the same in the IDM as in the SM. This also means that the renormalisation of the (extra) parameters of the IDM (masses of the scalars and the couplings $\lambda_L$ and $\lambda_2$) is not needed for these processes. All effects of the IDM will be indirect and will only affect the electroweak corrections. Because no extra coloured degree of freedom is brought by the IDM, the NLO QCD corrections in the IDM and the SM are the same. However, we will go through the different steps of the NLO calculations. This will also serve to highlight how our general approach to loop calculations with the help of {\tt SloopS} which we have developed in the previous section, in particular of the soft/collinear singularities, comes into play.

\subsection{NLO QCD corrections}
\label{NLO_QCD}
\subsubsection{QCD corrections for (direct) $qq$-initiated production}
\label{sec:subsec_NLO_QCD}
\begin{figure}[!htb]
\centering
\includegraphics[width=0.25\textwidth,height=0.25\textwidth]{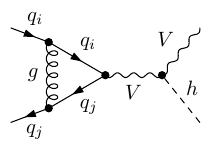} \qquad
\includegraphics[scale=0.9]{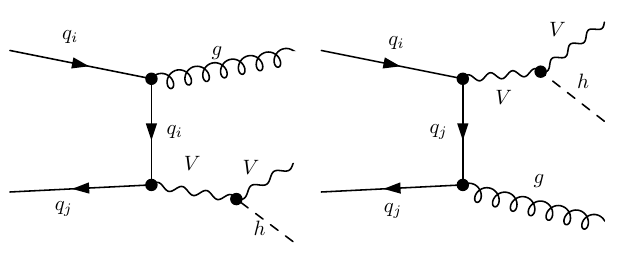}
\caption{Feynman diagrams for the virtual QCD correction (left) and real bremsstrahlung (right) for Higgs-strahlung. $V=W^\pm, Z$.}
\label{fig:qcdloop}
\end{figure}

The one-loop virtual QCD corrections consist essentially of the vertex correction shown in Fig.~\ref{fig:qcdloop}. 
In {\tt SloopS}, the two-point self-energy corrections on the external legs are combined with the corresponding counterterms. They are therefore not shown. 
The real corrections are shown in Fig.~\ref{fig:qcdloop} also. \\

In combining the virtual and real corrections we have checked the slicing method, $ \Delta \sigma^{\text{NLO}}_{\text{TCPSS}}$ Eq.~(\ref{TCPSS}), against the dipole subtraction $ \Delta \sigma^{\text{NLO}}_{\text{Dipole}}$ in Eq.~(\ref{dipole}). 
The result of the $qq$ initiated production $W^- h$ is shown in Fig.~\ref{fig::DS-TCPSSqcd}. From the middle panel of Fig.~\ref{fig::DS-TCPSSqcd}, the median results of TCPSS are within a precision much lower than $0.6\%$ for $\delta_s<10^{-4}$, when the results plateau. The results of DS have an even far better accuracy for the whole range of $\alpha^\prime$. 
$R_{T/D}$ which is a measure of the agreement between the two approaches, shows excellent agreement, again well below $0.6\%$ agreement for values of $\delta_s$ for which the results of TCPSS have stabilised. 
The comparisons clearly explain the efficacy of the dipole subtraction which provides a stable result for all values of $\alpha^\prime$. In the slicing method, convergence occurs for the smallest values of $\delta_s$, around $10^{-4}$, while smaller values give a larger Monte Carlo error; see Fig.~\ref{fig::DS-TCPSSqcd}. 
Note also that to reach this conclusion, for each $\delta_s$ (compared to $\alpha^\prime$) ten times more statistics were needed in the case of the TCPSS (particularly in the evaluation of the $\sigma_{H\bar{C}}$ part), leaving us without doubt preferring the power of the dipole approach.\\

\begin{figure}[!hbt]
\centering
\includegraphics[scale=0.6]{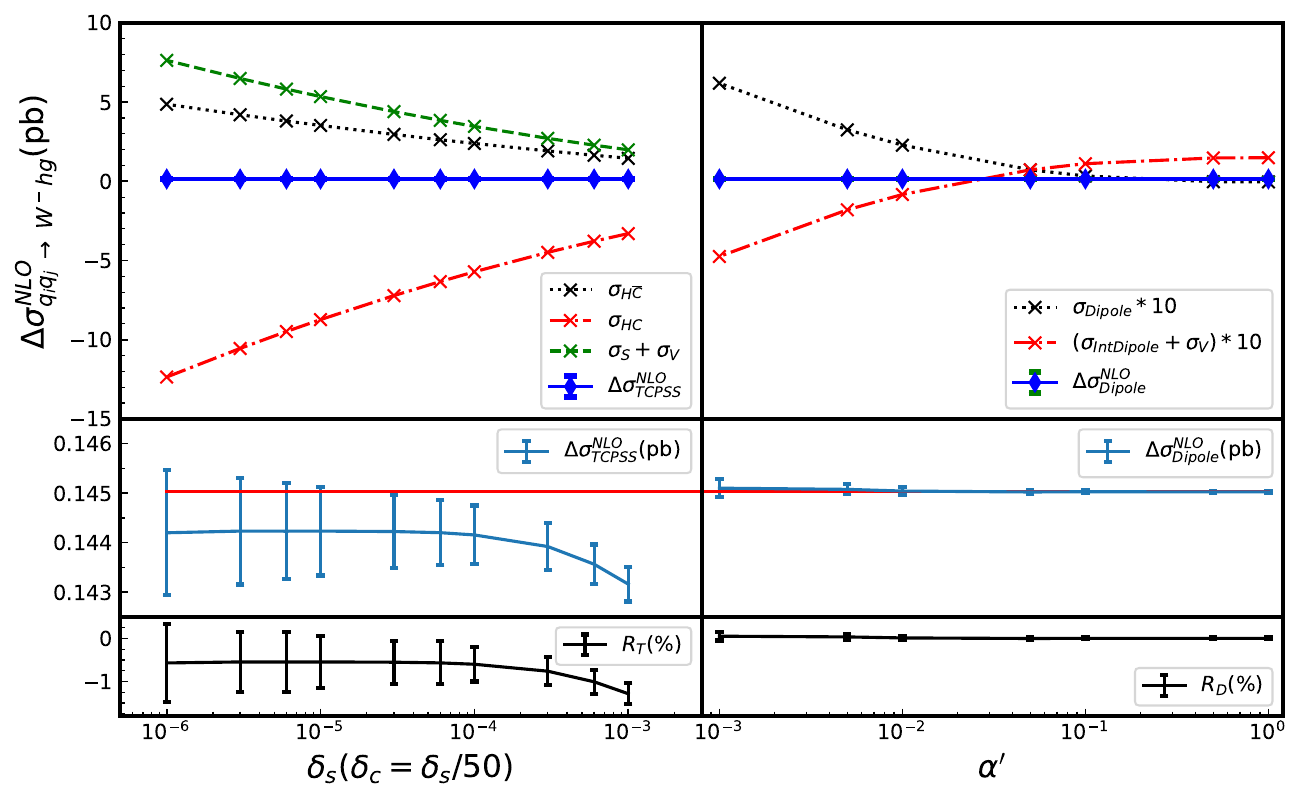}
\caption{Comparing the slicing method and dipole subtraction for QCD radiative corrections to $q_i q_j \to W^-hg$. 
In the upper panels, we show the results for each part for the DS and the TCPSS methods. 
The middle panels zoom in on the total contribution ($\sigma^{\text{NLO}}_{\text{TCPSS}}$  and $ \sigma^{\text{NLO}}_{\text{Dipole}}$)  of each method together with its Monte Carlo uncertainty. 
The solid line corresponds to the  central values returned by the Monte Carlo for each $\delta_s$, the (error) bars represent the result of the Monte Carlo uncertainty. 
Taking $ \sigma^{\text{NLO}}_{\text{Dipole}}(\alpha^\prime=1)$ as a reference, the lower panels show, in $\%$,  the corresponding  deviations from this reference point as $R_T=\frac{\Delta \sigma^{\text{NLO}}_{\text{TCPSS}}(\delta_s)-\Delta \sigma^{\text{NLO}}_{\text{Dipole}}(\alpha^\prime=1)}{\Delta \sigma^{\text{NLO}}_{\text{Dipole}}(\alpha^\prime=1)}$ and $ R_D=\frac{\Delta \sigma^{\text{NLO}}_{\text{Dipole}}(\alpha^\prime)-\Delta \sigma^{\text{NLO}}_{\text{Dipole}}(\alpha^\prime=1)}{\Delta\sigma^{\text{NLO}}_{\text{Dipole}}(\alpha^\prime=1)}$. 
For all but $\sigma_{H\bar{C}}$, the Monte Carlo is run with a statistics of $N=12\times 10^{6}$. 
For the $\sigma_{H\bar{C}}$ part, 10 times more statistics is required. The results are shown for $\sqrt{s}=13$ TeV and $\mu_R=\mu_F=2\;(M_V + M_h)$.}
\label{fig::DS-TCPSSqcd}
\end{figure}

\subsubsection{$gq$-induced real corrections}

These contributions are shown in Fig.~\ref{fig::qcdreal2}.
\begin{figure}[!th]
\centering
\includegraphics[scale=0.7]{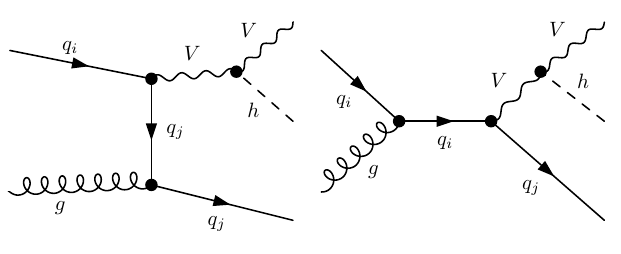}
\caption{Feynman diagrams of QCD (initiated) real radiation processes for  Higgs-strahlung.
}
\label{fig::qcdreal2}
\end{figure}

They feature a collinear divergence which is treated by the two methods we have just reviewed but in the absence of the virtual/soft contribution terms. As Fig.~\ref{fig::DS-TCPSSqcdqg} reveals, the plateau in $\delta_c$ for the TCPSS cut method is reached quickly. 
The dipole method is slightly more precise. The two methods agree better than the per-mil level.
\begin{figure}[H]
\centering
\includegraphics[scale=0.5]{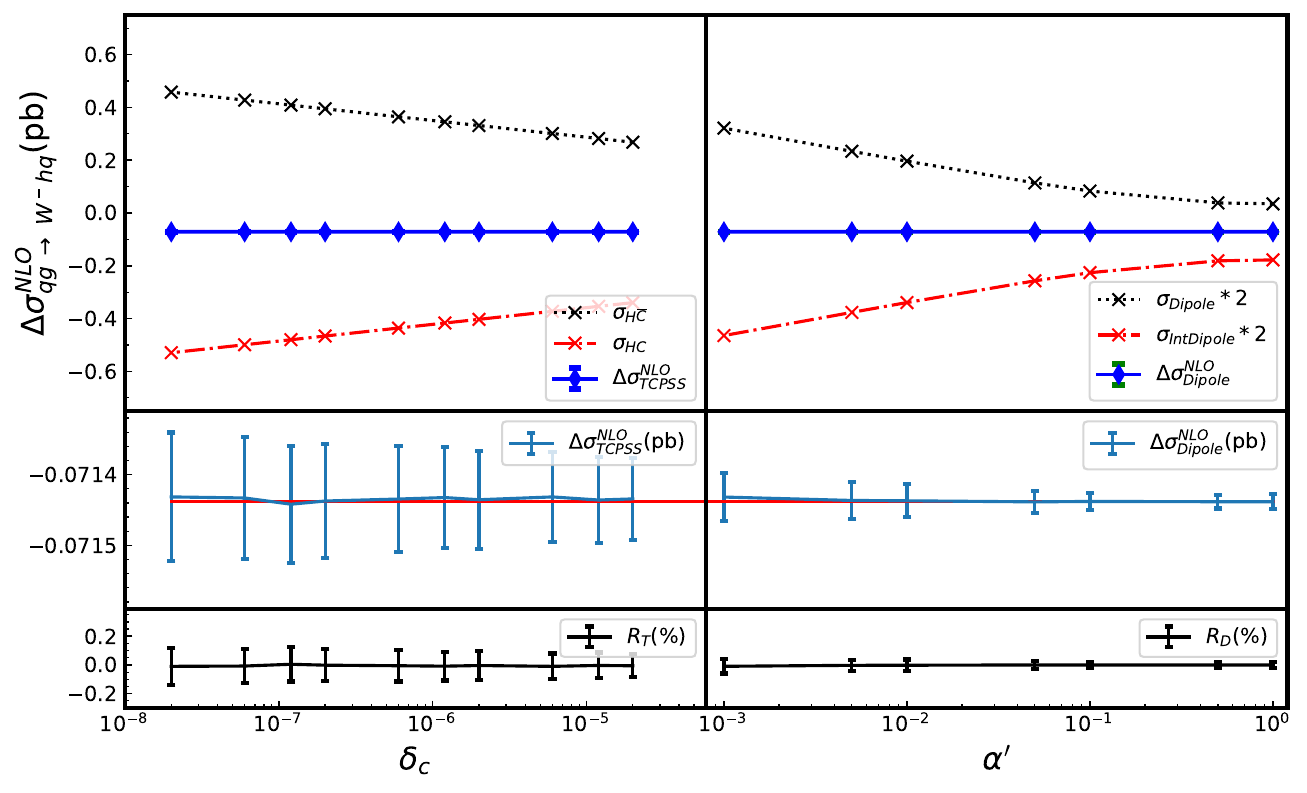}
\caption{As in Fig.~\ref{fig::DS-TCPSSqcd}.
Comparison of the cutting method and the dipole subtraction, but for contributions induced by $gq$. }
\label{fig::DS-TCPSSqcdqg}
\end{figure}

\subsubsection{$gg$-induced contributions: $Zh$ production}
$Zh$ production has an additional $gg \to Zh$ one-loop  (UV and IR finite) contribution, shown in Fig.~\ref{fig::ggzhsm}. The top plays an important role in this $gg$ fusion.

\begin{figure}[!ht]
\centering
\includegraphics[scale=1]{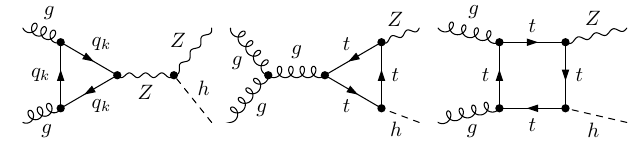}
\caption{The Feynman diagrams of $gg$ induced contributions for $Zh$ production. In the first diagram the contribution from the quarks of the first two generations (all massless) cancel. Only the top and bottom loops contribute, the vector part of $Z gg$ (real gluons) is zero as a consequence of Landau's theorem \cite{Landau:1948kw,Yang:1950rg}.
}
\label{fig::ggzhsm}
\end{figure}

\subsubsection{Comparison with {\tt Madgraph} for the NLO QCD corrections in the SM}

Taking all the QCD NLO contributions discussed in the previous section, we have conducted a tuned comparison (convoluting with the same PDF set and taking the SM parameters defined in Eq.~(\ref{paras-SM})
of the NLO QCD corrections between our code and {\tt Madgraph} \cite{Alwall:2014hca} in all three $Vh$ subprocesses. The numbers we show in Table~\ref{tab::mad-sloopsII} refer to our implementation of the dipole subtraction method (as discussed above, the 2-cut method gives exactly the same results). As can be seen in Table.~\ref{tab::mad-sloopsII} the agreement is excellent. The agreement is for the most part below the level of the per mil and the worst is at $0.4\%$. \\

\small
\begin{table}[!htb]
\begin{center}
\renewcommand\arraystretch{1.2}
\begin{tabular}{| p{1.5cm}<{\centering}  |p{2.5cm}<{\centering} |p{3cm}<{\centering}|p{3cm}<{\centering}|}
\cline{3-4}
\multicolumn{2}{c|}{}& Madgraph  & SloopS \\
%
\hline
\multirow{2}*{$Zh$} 
&LO &  0.6021(5)   & 0.6022(1) \\
\cline{2-4}
& NLO &   0.825(2)  & 0.824(1)\\
\hline
\multirow{2}*{$W^-h$}
&LO & 0.4282(4) &0.4283(1) \\
\cline{2-4}
&NLO(QCD) & 0.528(2)  & 0.5279(2) \\
\hline
\multirow{2}*{$W^+
h$}
&LO & 0.6803(6)   & 0.6804(2) \\
\cline{2-4}
&NLO(QCD) & 0.839(3) & 0.8420(4) \\
\hline
\end{tabular}
\end{center}
\caption{Comparison between the results of the LO and NLO QCD corrections in {\tt Madgraph} and {\tt SloopS} in the SM. Cross sections are in pb. For $Zh$, the one-loop $gg$ initiated process is included. Here we take the  {\tt NNPDF31-luxqed} PDF set. The renormalisation and factorisation scales are set to $\mu_R=\mu_F=M_Z$. }
\label{tab::mad-sloopsII}
\end{table}
\normalsize
\vspace*{-0.5cm}

\subsubsection{Scale and PDF uncertainties }

As mentioned earlier (see Sect.~\ref{sec::vhsm}), computations for the processes at hand have now been performed beyond NLO and beyond inclusive cross sections. 
The remaining uncertainties have been estimated. 
For the QCD contributions, these uncertainties are of the order of 1-2\% (see the LHC Higgs cross-section handbooks, \cite{LHCHiggsCrossSectionWorkingGroup:2011wcg,Dittmaier:2012vm,LHCHiggsCrossSectionWorkingGroup:2013rie,LHCHiggsCrossSectionWorkingGroup:2016ypw}). 
Estimating these uncertainties is important because they set sensitivity benchmarks, independent of experimental uncertainties, above which the effect of any new physics can be confirmed.
Based on the NLO corrections results of the same code that derives the contributions of the New Physics, we re-evaluate the uncertainties with the latest PDF. 
We will base our analyses on three PDF sets: {\it i}) {\tt PDF4LHC21\_40} \cite{PDF4LHCWorkingGroup:2022cjn} which is  a combination of the {\tt  CT18} PDF, the {\tt NNPDF3.1} \cite{NNPDF:2017mvq} and the {\tt MHST20} \cite{Bailey:2020ooq}
{\it ii}) {\tt NNPDF31-luxqed } (our default set) and {\it iii}) {\tt MSHT20qed\_nnlo} \cite{Cridge:2021pxm}. 
In both {\tt NNPDF31-luxqed } and  {\tt MSHT20qed\_nnlo} the photon content is implemented with some realisation of the {\tt LUX} approach \cite{Manohar:2016nzj,Manohar:2017eqh}. 
The latter sets permit for the incorporation of photon-initiated contributions, making it possible to take into account the effect of both the QCD corrections and the full NLO EW.  The {\tt LUX} approach has allowed a much better determination of the photon content of the proton. 
We use {\tt PDF4LHC21\_40} not only to weigh the importance of the photon-initiated contribution in our processes but also to check whether the precision of the fit including the photon PDF does not deteriorate much the overall uncertainty of the prediction due to the PDF.
One of the uncertainties in the calculation of the cross section arises from the scale variation. Recall that for the scale dependence we take $\mu_R=\mu_F$ and that we track the dependence through the parameter $x_{R/F}$ with $\mu_{R/F}=x_{R/F}\;M_{HV}= x_{R/F}\;(M_V+M_h)$. 
\begin{figure}[tb!]
\begin{center}
\includegraphics[height=0.47\textwidth,width=0.42\textwidth]{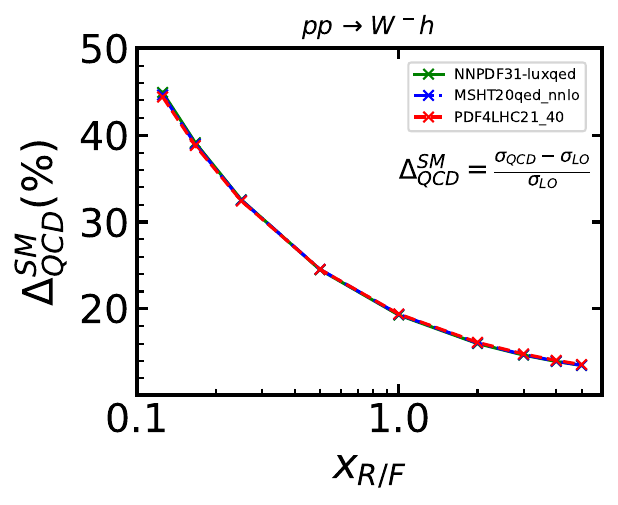} \hspace*{0.1\textwidth}
\includegraphics[height=0.47\textwidth,width=0.42\textwidth]{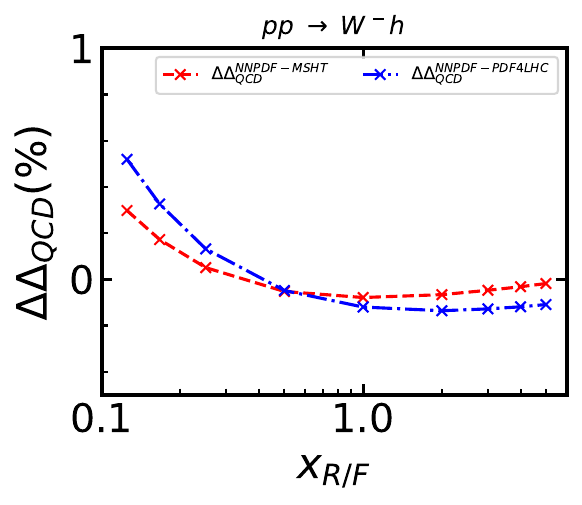}
\includegraphics[height=0.47\textwidth,width=0.42\textwidth]{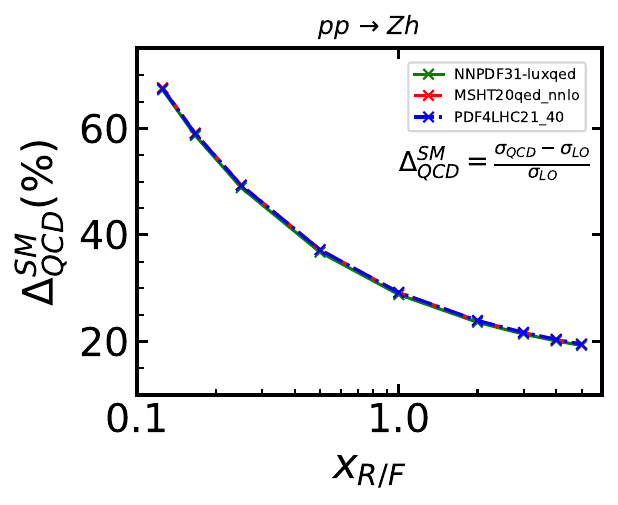} \hspace*{0.1\textwidth}
\includegraphics[height=0.47\textwidth,width=0.43\textwidth]{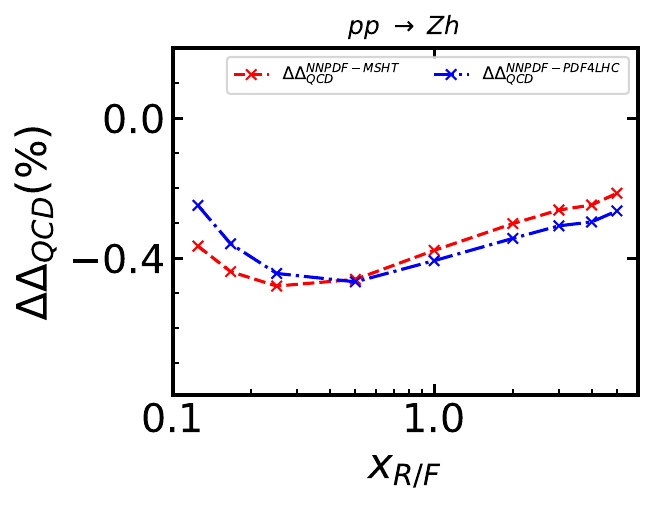}
\caption{The scale and PDF uncertainty of the NLO QCD (percentage) 
correction, $\Delta_{\text{QCD}}^{\text{SM}}$, for $pp \to V h$ process.
For better visibility of the error coming from the different choices of the PDF, we define the percentage differences (between the relative corrections) as $\Delta \Delta_{QCD}^{NNPDF-MSHT} = \Delta_{QCD}^{SM}(NNPDF) - \Delta_{QCD}^{SM}(MSHT)$ and $\Delta \Delta_{QCD}^{NNPDF-PDF4LHC} = \Delta_{QCD}^{SM}(NNPDF) - \Delta_{QCD}^{SM}(PDF4LHC) $. The Monte Carlo integration errors are below the per-mil level and are therefore not displayed in these figures. The results for $W^+h$ are not shown since they are  practically the same as for $W^-h$.
}
\label{fig::muqcd}
\end{center}
\end{figure}

As shown in Fig.~\ref{fig::muqcd},  we first recover the large NLO QCD corrections \cite{Han:1991ia,Brein:2003wg} and observe significant uncertainties, in particular the large corrections for $x_{R/F}<1$ (these are essentially the same as for the NLO QCD corrections to Drell-Yan) in all three channels. Observe that the NLO scale dependence almost plateaus when $x_{R/F}>1$. 
We will sometime quote  results for $x_{R/F}=2$, when considering the scale uncertainty. 
It just happens that our results with $x_{R/F}=2$  are consistent with the inclusion of the (small) NNLO and N$^3$LO \cite{Baglio:2022wzu} for the {\it inclusive} cross sections.
Fig.~\ref{fig::muqcd} also shows that the PDF uncertainty estimated, for this section,  as the difference between the three PDF sets we have chosen is at the per-mil level, in particular around the $x_{R/F}$ values we have suggested. This serves as further confirmation that our code is performing well. However, it should be noted that the estimated uncertainty could be greater if
$x_F$ and $x_R$ varied independently or if a specific distribution (including decays of $V$, $h$) were considered.

\subsection{EW corrections in the SM}

\subsubsection{Renormalization. The $G_\mu$ scheme}
The renormalisation of the IDM \cite{Banerjee:2021oxc} is carried out in the On-Shell (OS) scheme with a slight variation in relation to the renormalisation of the electromagnetic coupling constant $\alpha$. In the standard OS scheme \cite{Belanger:2003sd} the latter is defined in the Thomson limit (vanishing photon momentum transfer). Using the Ward identity and OS condition, the renormalization constant for the electric charge $\delta Z_e$ is obtained as
\begin{equation}
\begin{aligned}
\delta Z_e^{\alpha(0)} &= -\frac{1}{2}\delta Z_{AA} - \frac{1}{2}\frac{s_W}{c_W} \delta Z_{ZA} =\frac{1}{2}\frac{\partial \Sigma_T^{AA}(Q^2)}{\partial Q^2} \bigg|_{Q^2 \to 0} - \frac{s_W}{c_W} \frac{\Sigma_T^{AZ}(0)}{M_Z^2},
\end{aligned}
\end{equation}
where the superscript relates to the $\alpha(0)$-scheme. For the $A \to B$ transition,
$\delta Z_{AB}$ and $\Sigma_T^{AB}(Q^2)$ are, respectively, the renormalisation constants of the wave function and the transverse parts of the $A \to B$ self-energy at the squared momentum transfer $Q^2$. A subtraction at $Q^2=M_Z^2$ ( through the running fine-structure constant $\alpha(M_Z^2)$,  $\alpha(M_Z^2)$-scheme) would be more appropriate. In fact, we adopt the $G_\mu$-scheme with $G_\mu$, the Fermi constant as defined in $\mu$ decay, as an independent parameter instead of $\alpha(0)$. The corresponding effective electromagnetic coupling strength is defined as \cite{Sirlin:1980nh}

\begin{equation}
\begin{aligned}
\alpha_{G_\mu}=\frac{\sqrt{2}G_F M_W^2}{\pi}\left(1-\frac{M_W^2}{M_Z^2} \right)=\alpha(0)(1+\triangle r),
\end{aligned}
\end{equation}
with the corresponding counterterm for $\alpha$ 
\begin{equation}
\begin{aligned}
\delta Z_e^{G_\mu} = & \delta Z_e^{\alpha(0)} -\frac{1}{2}\triangle r.
\end{aligned}
\end{equation}
In the usual linear gauge, with 
the Feynman parameter $\xi=1$, $ \triangle r$ is expressed as~\cite{Sirlin:1980nh,Denner:1991kt} \begin{equation}\label{eq::deltar}\begin{aligned}
\triangle r =&\frac{\partial \Sigma^{AA}(Q^2)}{\partial Q^2}\bigg\vert_{Q^2=0}-\frac{c_W^2}{s_W^2}\left(\frac{\Sigma^{ZZ}_T(M_Z^2)}{M_Z^2} -\frac{\Sigma_T^W(M_W^2)}{M_W^2}  \right) +  \frac{\Sigma_T^{WW}(0)-\Sigma_T^{WW}(M_W^2)}{M_W^2} \\
&+2\frac{c_W}{s_W}\frac{\Sigma_T^{AZ}(0)}{M_Z^2} + \frac{\alpha}{4\pi s_W^2}\left(6+\frac{7-4s_W^2}{2s_W^2} \log{c_W^2}\right).
\end{aligned}
\end{equation}

In the general implementation of a New Physics model and its renormalisation, in a first stage we check that renormalisation is carried correctly in the $\alpha(0)$ scheme by testing the UV finiteness and the gauge parameter independence of a series of processes. The latter exploits a multi-parameter  gauge-fixing function in the non-linear gauge. In the context of the IDM, the reader should refer to \cite{Banerjee:2021anv}. Only then, for LHC applications, do we revert to $\alpha_{G_\mu}$ scheme. Note also that for this class of models, the contribution to $\mu$ decay is only to two-point functions whose contributions to $S,T$ have also been independently evaluated for the parameter space of the IDM to pass the constraints; see Sect.~\ref{subsub::exp_constraints}.

\subsubsection{Virtual one-loop  electroweak corrections}
\label{sub:EWvirtual}
\begin{figure}[H]
\begin{center}
\includegraphics[width=\textwidth,height=10cm]
{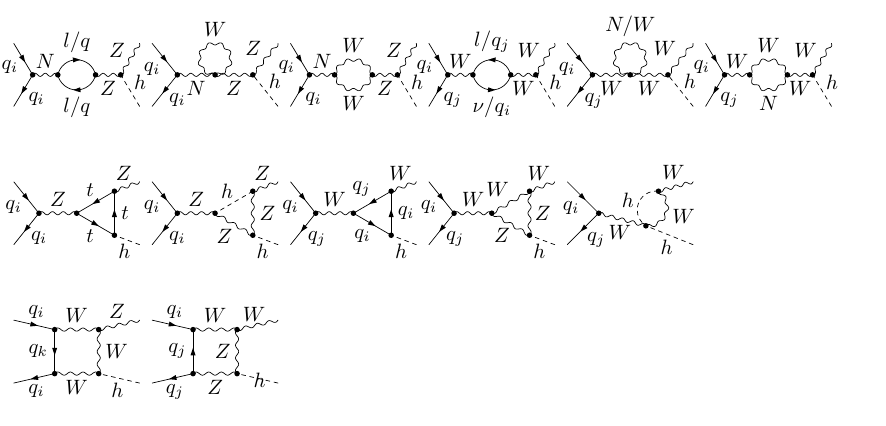}
\caption{Representatives of the set of EW virtual corrections for $q \bar q$ annihilation for $Wh$ and $Zh$ production. The first row contains the self-energy diagrams, the second the triangles, and the third is a collection of boxes. $N=Z,\gamma$. }
\label{fig::ewloop}
\end{center}
\end{figure}

Compared to the QCD corrections, the full set of EW virtual contributions is much larger. It contains 2-point, 3-point, and 4-point one-loop diagrams, some of which are displayed in Fig.~\ref{fig::ewloop}.

Another set is displayed in Fig.~\ref{fig::ewloopqed}. The latter (with photonic contributions) exhibits infrared (soft and collinear) singularities that must be combined with the real corrections.

\begin{figure}[H]
\centering
\includegraphics[scale=0.9]{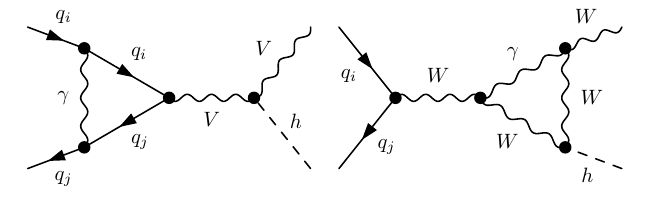}
\caption{Representatives of the virtual electroweak corrections with IR (photonic) divergences. 
The diagram on the left is the photon counterpart of the  QCD virtual gluon exchange in the first diagram in Fig.~\ref{fig:qcdloop}.} 
\label{fig::ewloopqed}
\end{figure}

\subsubsection{Real photonic corrections for $q \bar q$ annihilation}
\label{sub:EWreal}
The Feynman diagrams for EW real corrections for Higgs-strahlung are shown in Fig.~\ref{fig::ewreal}.
Compared with the real (initial state) QCD corrections, there are extra contributions with final-state photon radiations for $W^\pm h$ productions. 
\begin{figure}[!ht]
\centering
\includegraphics[scale=0.8]{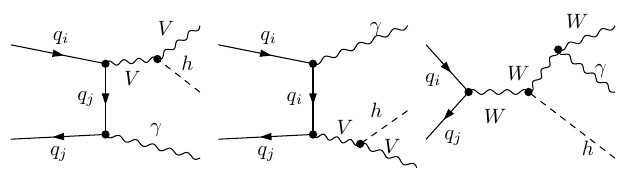}
\caption{Representatives of the real electroweak corrections with IR divergences. 
The first two diagrams combine with the first diagram of Fig.~\ref{fig::ewloopqed}. Photon emission from the $W$ combines with the second diagram of Fig.~\ref{fig::ewloopqed}.} 
\label{fig::ewreal}
\end{figure}
For  the radiation} of initial photons  in Fig.~\ref{fig::ewreal}, the subtraction of IR divergences is similar to the QCD case; see Sect.~\ref{sec:subsec_NLO_QCD} for the TCPSS and DS schemes.
We only need to substitute the electric charge for the colour factors and the strong coupling.
However, $W^\pm h$ productions involve contributions of final-photon radiation $W^\pm \to W^\pm + \gamma$. The latter features an extra IR soft divergence (but not a collinear one because of the large $W$ mass).

\subsubsection{Combining (photonic) real and virtual corrections soft/collinear divergences}
\label{sub:EWcombiqq}

\begin{figure}[!ht]
\center
\includegraphics[scale=0.65]{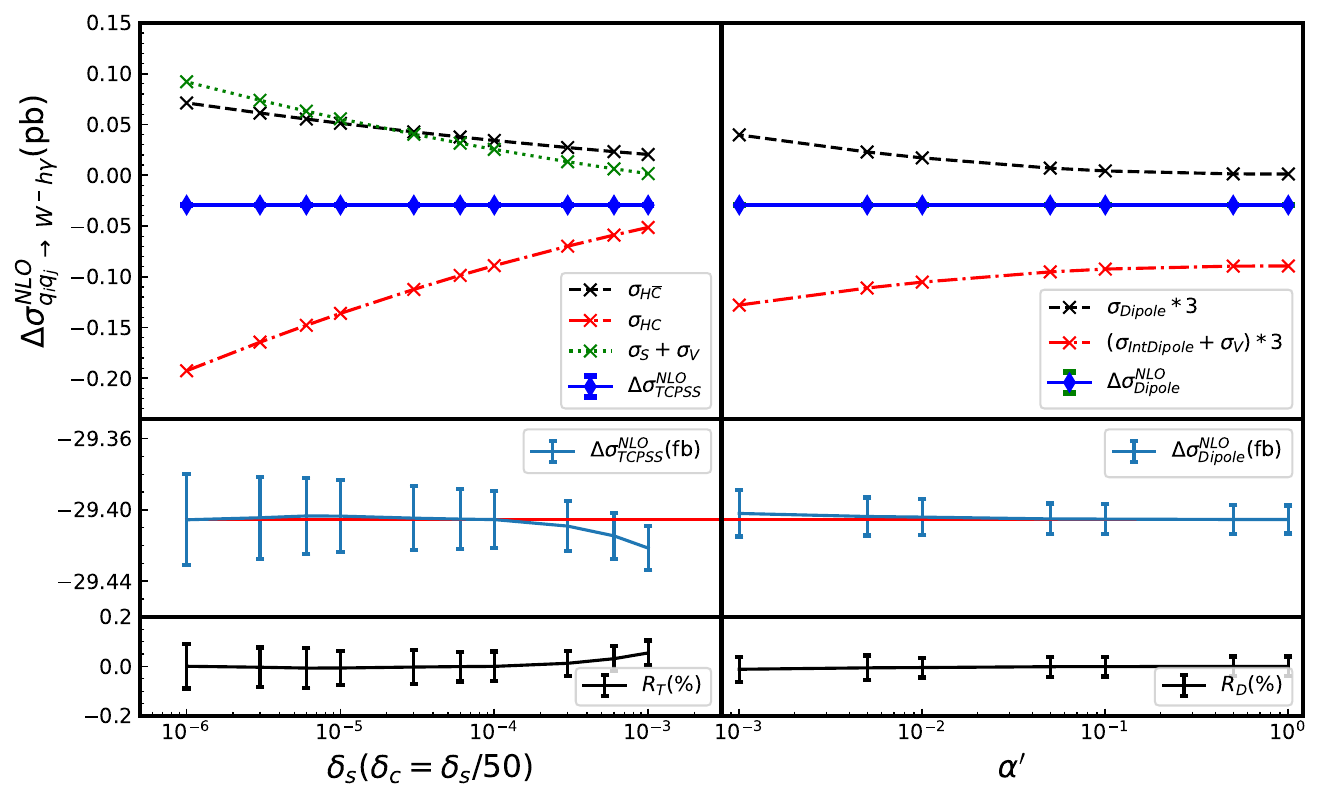}
\caption{As in Fig.~\ref{fig::DS-TCPSSqcd}
but comparing the slicing method and dipole subtraction for the photonic corrections. $R_{T,D}$ have the same meaning as in
Fig.~\ref{fig::DS-TCPSSqcd}.}
\label{fig::DS-TCPSSewqq}
\end{figure}
The combination of the virtual corrections from Sect. 3.8.2 and the real contributions from Sect. 3.8.3 follows a process similar to the steps outlined in Sect. 3.7.

Again, we compare dipole {\it versus} TCPSS in the electroweak case. As expected, there is good agreement. The dipole is not only more precise, but again TCPSS starts converging for extremely small values of the soft photon cut ($\delta_s$), while for extremely small $\delta_S$ the error gets larger. Nevertheless, the contribution is easily computed with an error less than the per-mil, see Fig.~\ref{fig::DS-TCPSSewqq}.
\subsubsection{$q \gamma$-initiated annihilation}
\label{sub:EWinduced}
\begin{figure}[H]
\centering
\includegraphics[width=\textwidth,height=2.7cm]{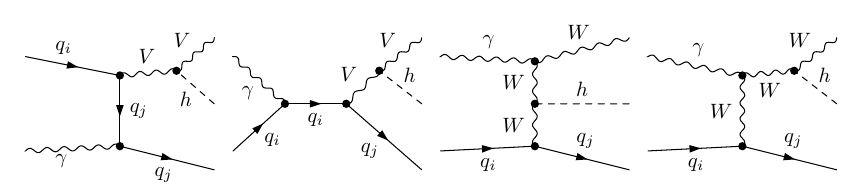}
\caption{Contributions to induced $\gamma q$ annihilation. Observe the presence of the $W$ in the $t-$channel for the for $Wh+q_j$ production.}
\label{fig::realqa}
\end{figure}
The application of DS and TCPSS schemes to $q \gamma$ annihilation is totally similar to the $q g$ case for the QCD corrections. 
However, a glance at the partonic contribution in the case of the photon reveals an important difference. 
While quark exchange, see Fig.~\ref{fig::realqa}, has an equivalent in the QCD case, observe the contribution of the $t-$channel spin-1 exchange (the $W$) for $W h q_j$ production. 
\begin{figure}[!bht]
\centering
\includegraphics[scale=0.57]{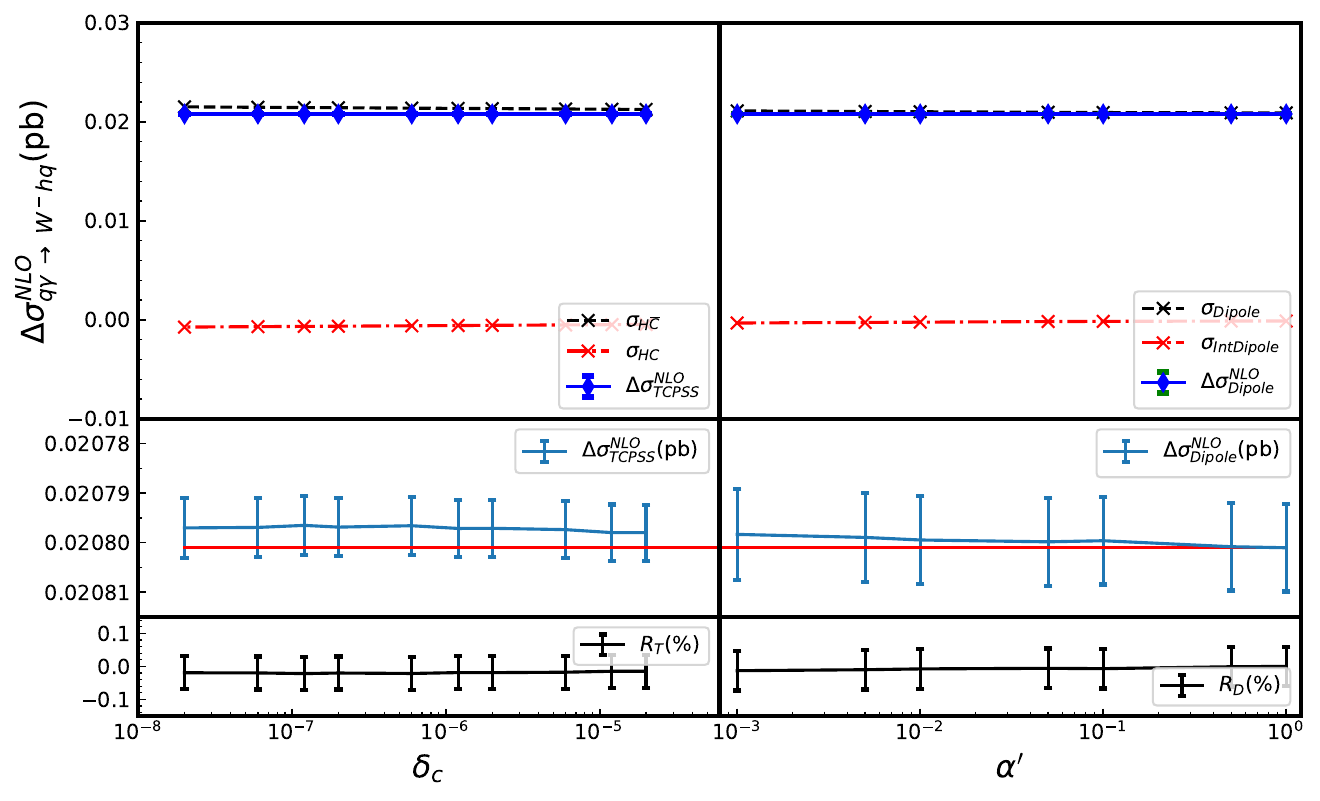}
\caption{Comparing the slicing method and dipole subtraction, but for induced $q\gamma$ annihilation contributions. 
}
\label{fig::DS-TCPSSewqa}
\end{figure}
For total cross sections, $t-$channel spin-1 exchange leads to large and constant ($\hat{s}$  independent) cross sections. 
Expect therefore that a large contribution will show up in the $Wh$ channel compared to $Zh$, for which such spin-1 exchange is absent. 
Note also that the large partonic cross section in the $W h q_j$ channel is due to the forward scattering region and that a $p_T$ cut will reduce this contribution. 
At this point, let us just note that the application of DS and TCPSS shows, see Fig.~\ref{fig::DS-TCPSSewqa}, the excellent agreement between the two approaches. 
They both show excellent stability over 3 orders of magnitude in the respective cut-offs. 

\subsubsection{$\gamma \gamma$-induced $Zh$ production in the SM}
\label{sub:EWgamgam}
\begin{figure}[H]
\centering
\includegraphics[scale=1.1]{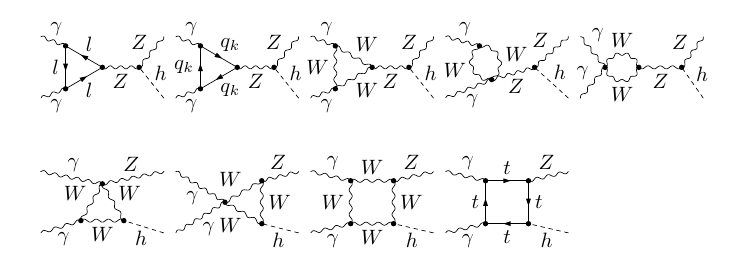}
\caption{Induced $\gamma \gamma$ contributions in the case of $Zh$ production.}
\label{fig::ewaazhsm}
\end{figure}

For $Zh$ production, there also are contributions from the $\gamma \gamma$-induced processes as shown in Fig.~\ref{fig::ewaazhsm}.
These contributions are totally negligible not only because these are generated at one-loop but also because the $\gamma \gamma$ luminosity (convolution) is extremely tiny.  In the end, this cross section amounts to less than $3 \times 10^{-6}$ of the total $Zh$ production at the LHC. We will not mention it in the rest of this paper. 

\subsubsection{The electroweak corrections in the SM: results, scale dependence and PDF uncertainty}
\begin{figure}[!h]
\centering
\setlength{\abovecaptionskip}{0pt}
\includegraphics[scale=0.645]{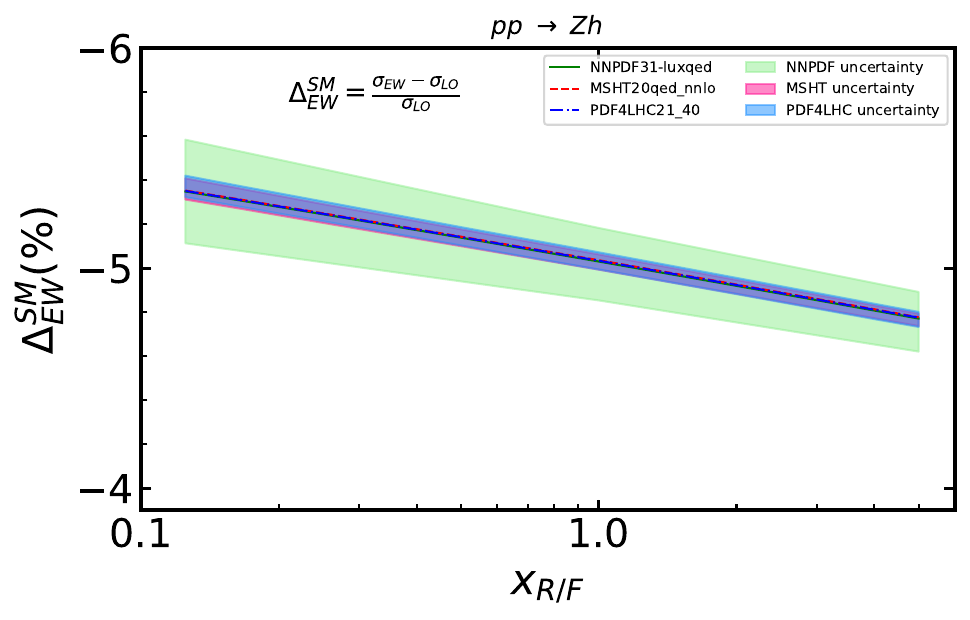}
\includegraphics[scale=0.645]{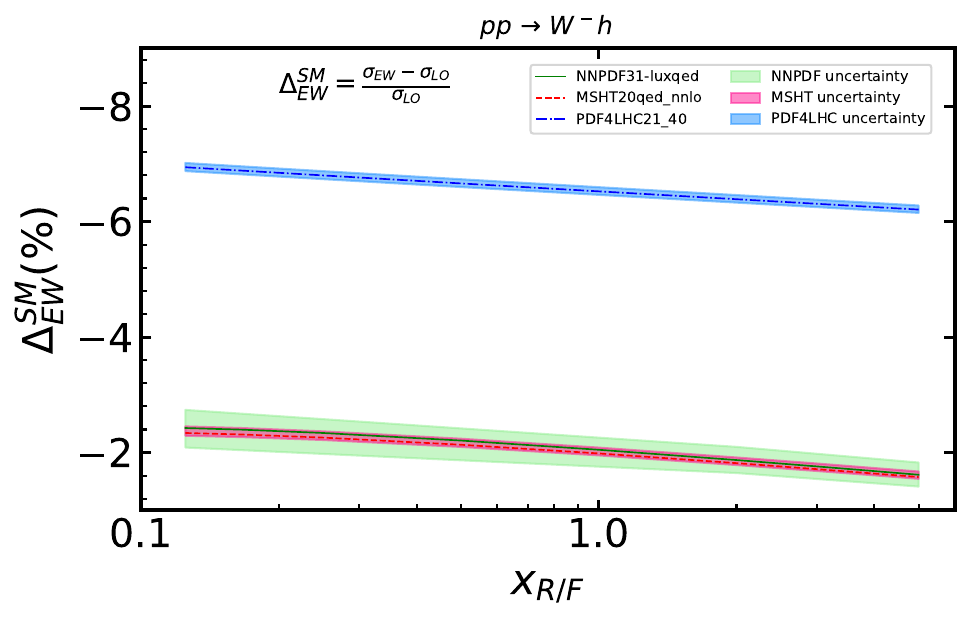}
\includegraphics[scale=0.645]{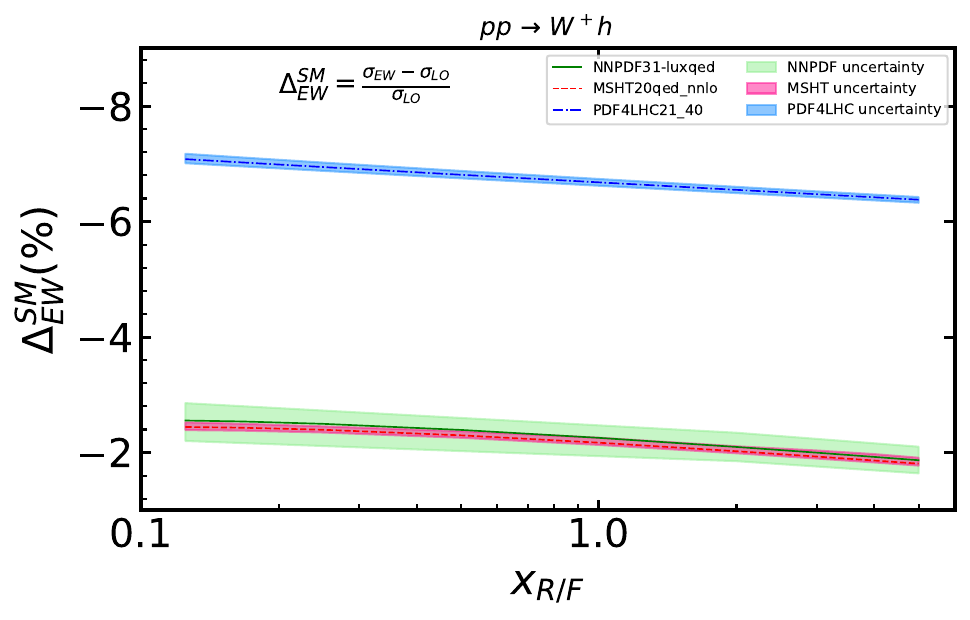}
\caption{The relative electroweak correction for the three $Vh$ channels, as a function of the scale variation parameterised through the variable $x_{R/F}$. Three sets of PDF are considered as discussed in the text.} 
\label{fig::muewpdf}
\end{figure}

Taking all the electroweak contributions, virtual corrections in ~\ref{sub:EWvirtual}, the real direct corrections in~\ref{sub:EWreal} and the $q \gamma$-induced contributions in~\ref{sub:EWinduced} (the $\gamma \gamma$ induced contributions in~\ref{sub:EWgamgam} are far too tiny),  our numerical results for the EW corrections in the SM for the inclusive $Vh$ cross sections at the 13 TeV LHC are shown in Fig.~\ref{fig::muewpdf} for our three sets of PDFs. 
The band, for each PDF set,  represents the precision due to the PDF uncertainty given by each PDF set. 

First, leaving aside the issue of the PDF uncertainty, the important lesson is that in contrast to the NLO QCD corrections, see Fig.~\ref{fig::muewpdf}, the scale uncertainty for the electroweak corrections is comparatively very small. The results of the corrections are rather stable in the range $1/8<x_{R/F}<5$.
 
For the $Zh$ channel, the results of the convolution with all three PDFs agree extremely well. The PDF uncertainty within PDF4LHC21 is the smallest. 
With about $-5\%$ relative correction for $Zh$,  we recover here the results of \cite{Ciccolini:2003jy} for $M_h=125$ GeV.

For $W^\pm h$, the electroweak corrections show important differences, depending on which PDF set is used. The differences extend beyond the inherent (very tiny) PDF uncertainty within each set. It is the most precise (in terms of the uncertainty given by the PDF set), {\tt PDF4LHC21\_40} that  disagrees with both {\tt NNPDF31-luxqed} and {\tt MSHT20qed\_nnlo}, the latter two agreeing extremely well with each other. This is due to the photon-induced process, whose contribution accounts for almost $5\%$ difference. Not fitting the photon content data may give a more precise PDF set but does not permit one to take into account some important subprocesses. 
As noted previously, the photon-induced contribution is tiny in the $Zh$ channel but not so in the $Wh$ channel. As we have previously underlined, the spin-1 $W$ $t$-channel exchange in $\gamma q \to Wh q^\prime$ is large. We also underline that our results agree extremely well with those of the original calculation of the electroweak corrections in $Vh$ conducted in \cite{Ciccolini:2003jy} when the induced photon contribution is switched off.  Considering that the agreement between the results of convoluting with these 3 PDF sets was at per-mil level for the QCD correction, the discrepancy in the NLO electroweak correction is essentially due to the contribution of the parameterisation (or lack of)  the photon content as seen in Fig.~\ref{fig::muewpdf}. 
To quantitatively substantiate the argument made earlier that a cut of $p_T^W$ would reduce the contribution of the induced photon cross section, $\sigma_{\gamma-\text{induced}}$, Table~\ref{tab:siggammaqrel} gives the relative size of this contribution (with cuts of {\tt NNPDF31-luxqed}) with increasing cuts on $p_T^W$. We observe that as the $p_T^W$ cut increases, this contribution decreases until it becomes almost negligible. This confirms our argument.
This may also explain why $\sigma_{\gamma-\text{induced}}$ in $Wh$ has been found to be on the order of percent in~\cite{Denner:2011id,Denner:2014cla}, albeit with an older set of PDF. \\

\begin{table}[!htb]
\renewcommand\arraystretch{1.5}
\begin{center}
\begin{tabular}{| p{1.8cm}<{\centering} |  p{1.8cm}<{\centering}| p{2.7cm}<{\centering} |p{2.7cm}<{\centering} |p{2.7cm}<{\centering} |}
\cline{2-5}
\multicolumn{1}{c|}{}& no cut & $P_T^W>20$ GeV & $P_T^W>100$ GeV & $P_T^W>200$ GeV  \\
\hline
$W^-h$   & 4.5\% & $ 4.5\%$ & $ 2.3\%$ & $ 0.9\%$  \\
\hline
$W^+h$   & 4.5\% &  4.5\% & 2.3\%   &  1\% \\
\hline
\end{tabular}
\end{center}
\caption{$\sigma_{\gamma-\text{induced}}/\sigma_{\text{LO}}$ for different values of the cut on $p_T^W$ for $W^\pm h$ production. The {\tt NNPDF31-luxqed} set is used for the LHC at 13 TeV.}
\label{tab:siggammaqrel}
\end{table}

\subsection{Electroweak and QCD NLO corrections in the SM combined}
\begin{figure}[!ht]
\includegraphics[scale=0.7]{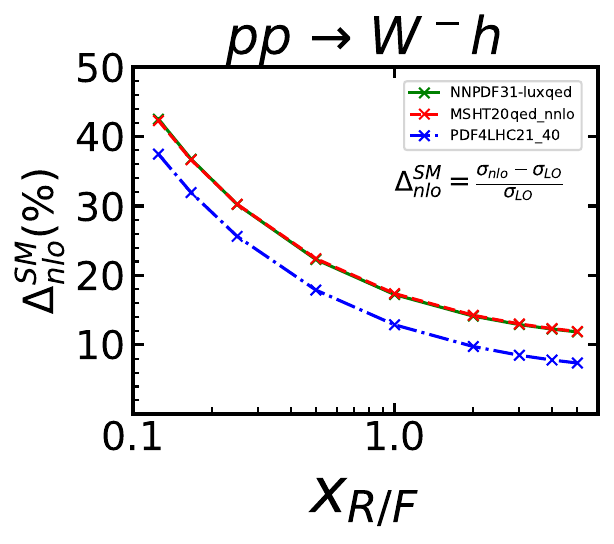}
\includegraphics[scale=0.7]{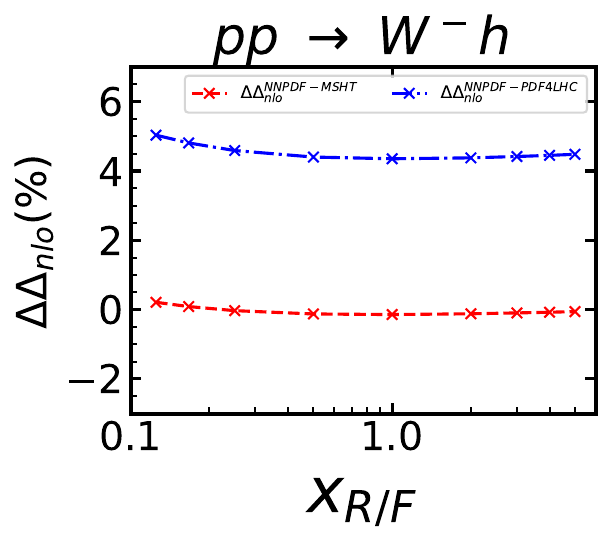}
\includegraphics[scale=0.7]{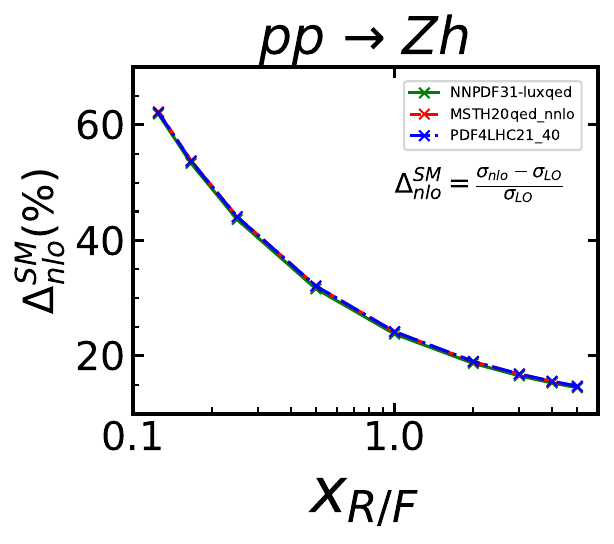}
\includegraphics[scale=0.7]{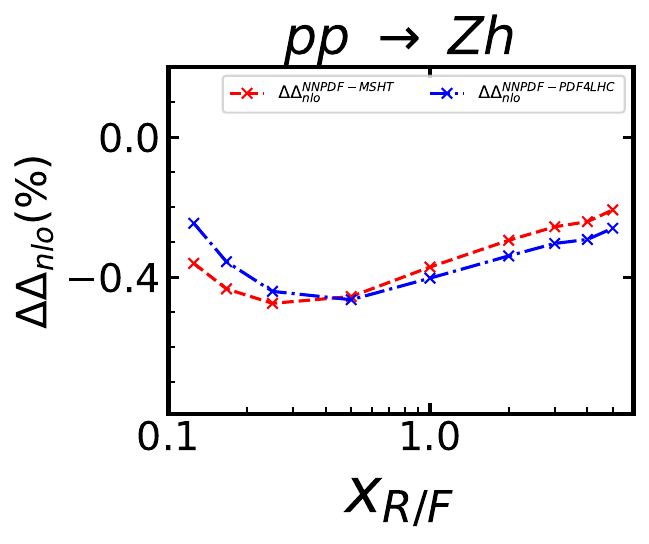}
\caption{As in Fig.~\ref{fig::muqcd} but for the full NLO, combined QCD and EW corrections for $Vh$. The panels on the right gives the difference in the percentage NLO correction taking the {\tt NNPDF-lux} as the default PDF set.}
\label{fig::munloSM}
\end{figure}
For the full one-loop corrections in the SM we add (linearly) the QCD and the electroweak corrections and convolute with three PDF sets. The results are displayed in Fig.~\ref{fig::munloSM}. This combination reveals the large-scale uncertainty in the wide range of $x_{R/F}$ we have considered, present in all three channels independently of the PDF set chosen. This large-scale uncertainty is due to the QCD NLO contribution. However, the differences in the predictions between the three sets of PDF are due to the incorporation or not of the $\gamma q$ induced process in the electroweak corrections. 
These differences are exhibited in the $Wh$ channel while the agreement is excellent (per-mil level) in the $Zh$ channel.

Before examining the impact of the IDM in these channels, it is crucial to emphasise that any potential deviation from the SM prediction should be analysed considering the theoretical uncertainties discussed and displayed in Fig.~\ref{fig::munloSM}. Our analysis is based on the NLO results (and LO for $gg \to Zh$). Furthermore, taking into account the recent analysis based on N$^3$LO\cite{Baglio:2022wzu}, which provides estimates for theoretical uncertainties (including scale variation, PDF, and their correlations), as well as results from higher-order analyses for $gg \to Zh$ \cite{Davies_2021,Chen_2021,Alasfar_2021,Wang_2022,Chen_2022}, the current benchmark for QCD accuracy (for inclusive cross sections) is just below the uncertainty of 2\% for $Wh$. We will assume a similar  2\% uncertainty for $Zh$.  PDF uncertainties should be reduced in the future, particularly by improving the extraction of N$^3$LO PDF along with their photon component. In this context, any deviation from the SM prediction of these cross sections should be considered including the potential indirect effects of the IDM, which we now address. Only when these SM theoretical uncertainties (missing higher orders, scale, and PDF uncertainties) are considered can we entertain, as we will do in due course, the experimental observability of the indirect effects of the IDM in these processes.

\section{Higgs-strahlung in the IDM: One loop results and deviations from the SM}
\label{sec::hv_idm}
\begin{figure}[!ht]
\centering
\includegraphics[scale=0.8]{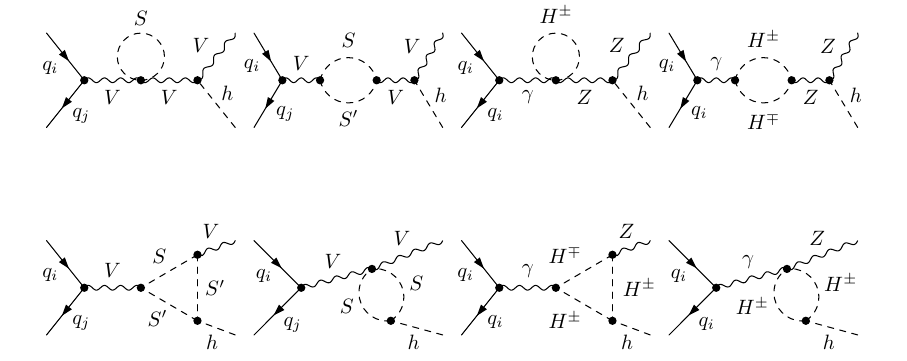}
\caption{Additional contributions from the IDM for $Vh$ production, $V = Z,W^\pm$. $S=H^\pm,X,A$. $Zh $ production has  additional contributions from $\gamma-Z$  mixing compared to $W^\pm h$ production. Diagrams with Goldstone bosons are not shown in this list.}
\label{fig::loopidm}
\end{figure}

The IDM does not change any of the couplings that enter the tree-level amplitudes for $Vh$ production, and any effect is therefore indirect and appears beyond tree-level. The IDM does not
add any new coloured particle, and the QCD corrections (virtual and real) are exactly the same as in the SM. The electroweak corrections for $pp \to Vh$, on the other hand, are sensitive to the virtual loop effects of the new scalars (in two-point and three-point functions), without affecting the infra-red/collinear structure. The complete set of the electroweak corrections due to the IDM is displayed in Fig.~\ref{fig::loopidm}. 
                                                                                                
Note that the contribution of the new scalars also enters in the renormalisation of the electroweak sector. As an example, the two-point functions that enter the definition of $\Delta r$ Eq.~(\ref{eq::deltar}) also now include the contributions of the new scalars. In this respect, it is important to recall that the $S,T$ parameters also constrain the contributions of the  two-point functions  involved in $Vh$ production. A variation of the three-point functions is also constrained by $h \to \gamma \gamma$. This is the reason why it is worth investigating how much additional sensitivity $pp \to Vh$ production has on the IDM parameters once all constraints are imposed.

\begin{figure}[!ht]
\centering
\setlength{\abovecaptionskip}{-0.5cm}
\includegraphics[scale=0.9]{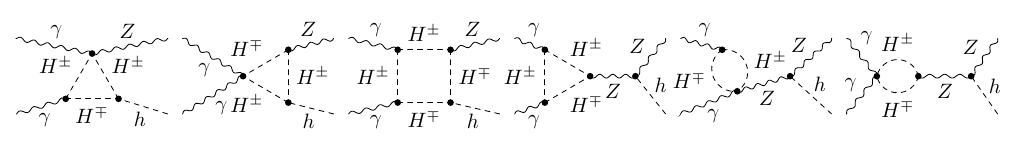}
\caption{Feynman diagrams of $\gamma \gamma \to  Zh$ including the contributions of charged Higgs.}
\label{fig::aazhidm}
\end{figure}

Potentially, there are also new contributions of the charged scalars of the IDM from $\gamma \gamma$-induced processes, as depicted in Fig.~\ref{fig::aazhidm}.
We find that these $\gamma \gamma$-induced contributions are totally negligible and are, in fact, much smaller than the SM $\gamma \gamma$ induced contributions of the SM, which are themselves too tiny. 

At the end of the day the departure from the SM prediction amounts to the discrepancy in the electroweak virtual corrections to the $q \bar q$ subprocesses between the IDM and the SM. We quantify the percentage change from the IDM through the variable $\delta$,                                        
\begin{equation}\label{eq::deltaidm}
\delta = \frac{\sigma_{IDM}^{NLO}-\sigma_{SM}^{NLO}}{\sigma_{SM}^{LO}}= \frac{\sigma_{IDM}^{virt-EW}-\sigma_{SM}^{virt-EW}}{\sigma_{SM}^{LO}} .
\end{equation}

\subsection{Scenarios without DM constraints}\label{sec:hv_idm_nodm}

\begin{figure}[!btp]
\centering
\includegraphics[height=6.5cm, width=0.48\textwidth]{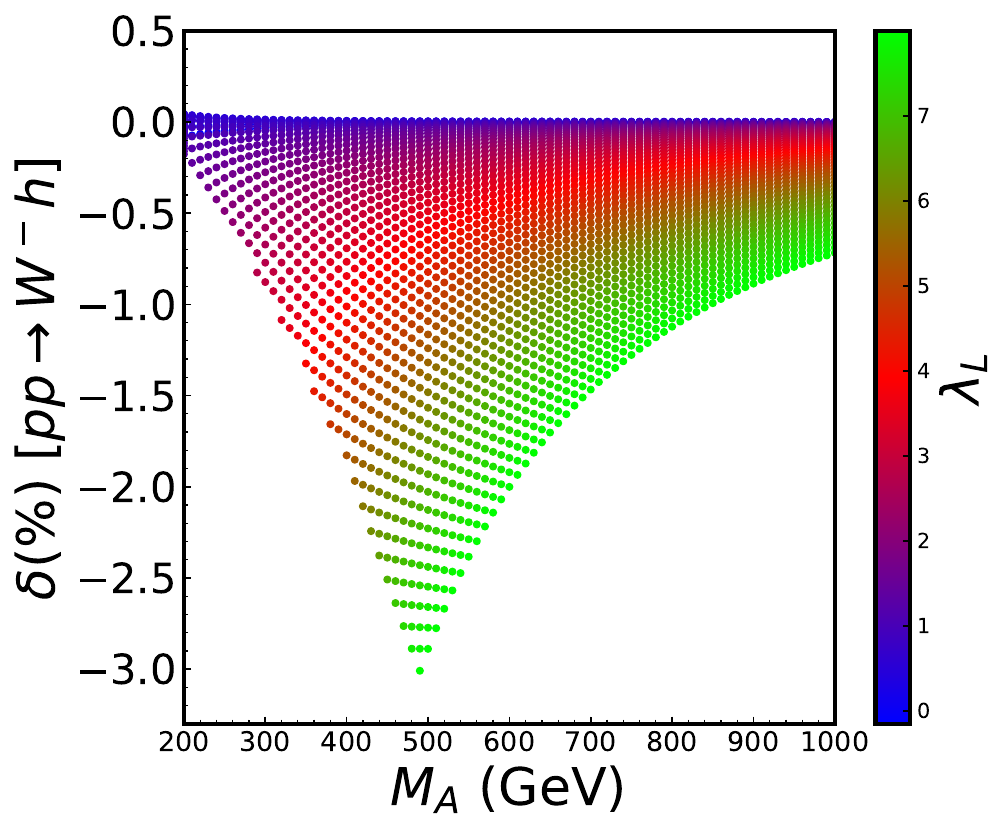}
\includegraphics[height=6.5cm, width=0.48\textwidth]{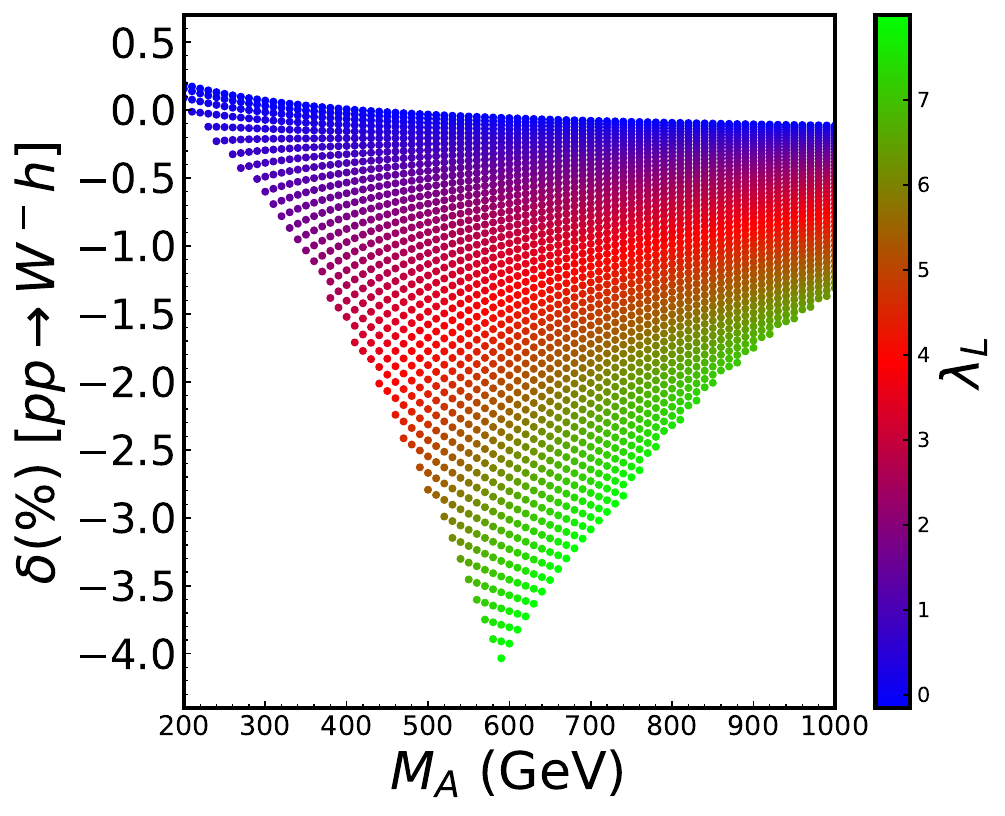}
\includegraphics[height=6.5cm, width=0.48\textwidth]{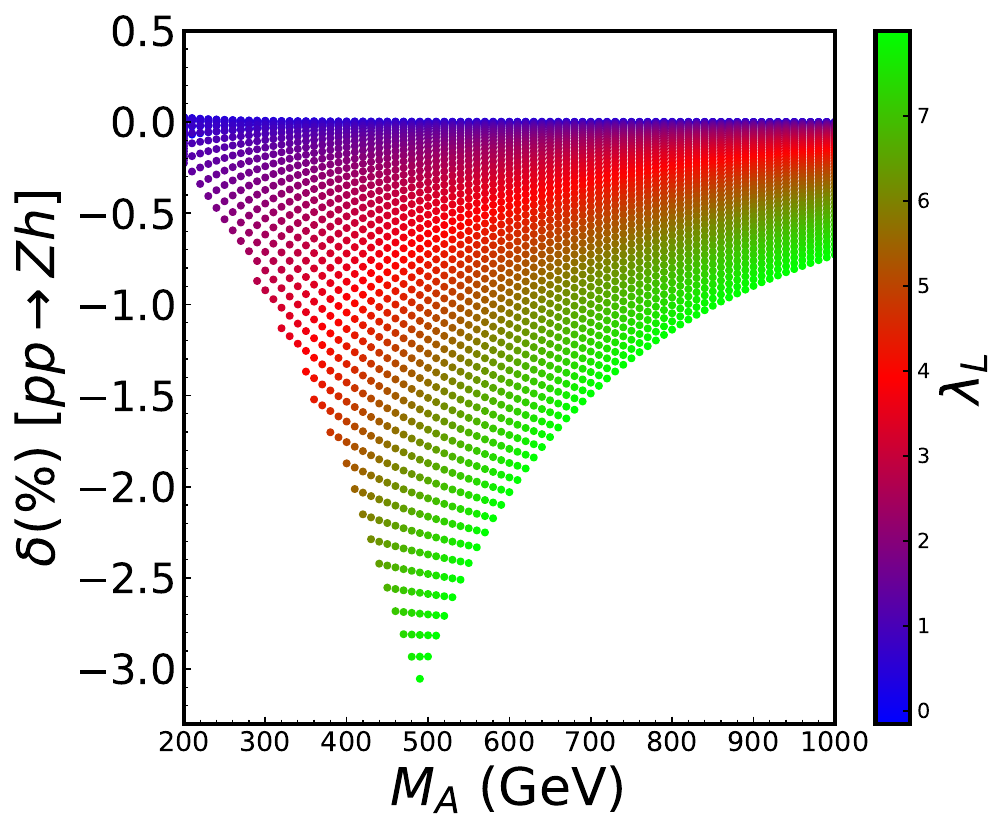}
\includegraphics[height=6.5cm, width=0.48\textwidth]{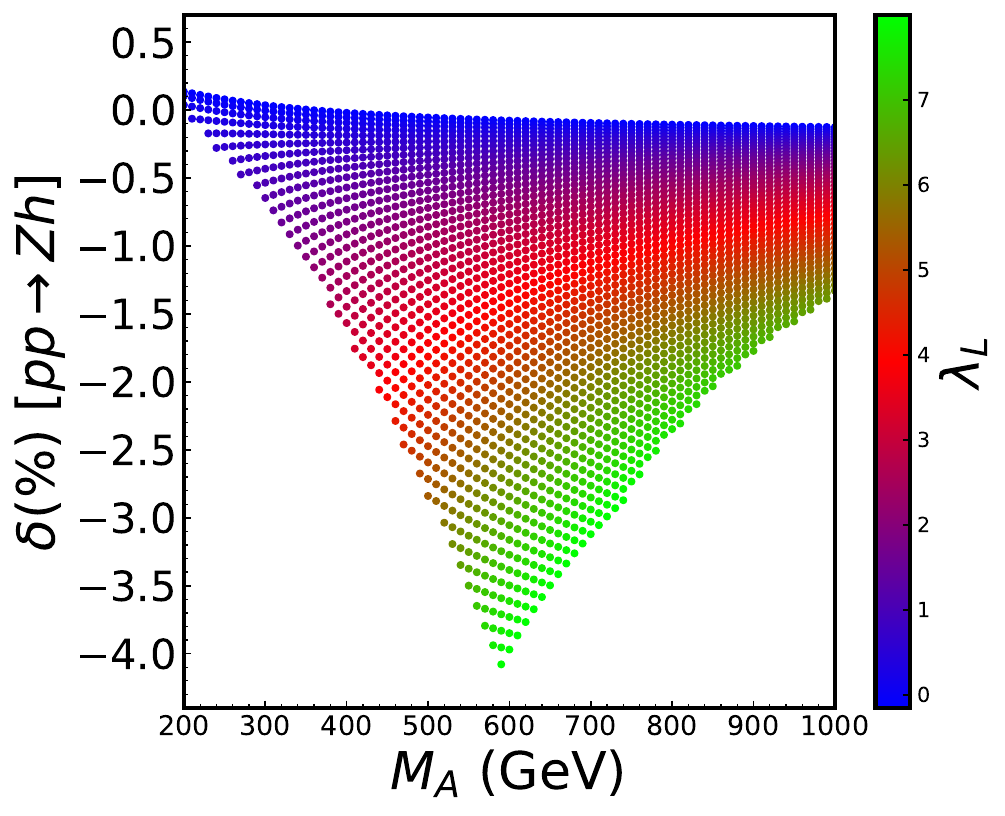}
\caption{The percentage difference between the SM and the IDM for $pp \to Vh$ cross sections when no DM constraints are imposed but all other constraints are taken into account (see Sect.~\ref{sec::currentconstraints}). 
In the panel on the left there is full degeneracy on the masses of the three extra scalars of the IDM: $M_X=M_A=M_{H^\pm}$ (models {\bf A}, see text). 
The range $(100<M_{A}<200)$ GeV is not displayed in this case, in accordance with the range in the panels on the right and also because this range leads to too small corrections. 
In the panel on the right we have $M_A=M_{H^\pm}$ (models {\bf B}) the mass of the lightest (neutral scalar) $X$ is taken such that $\Delta M =M_A-M_X = 100$ GeV. 
Note that the deviation $\delta$ for $W^+ h$ is exactly the same as that of $W^- h$ and is therefore not shown. 
BP0 is not representative of these two scenarios. 
BP1 belongs to the panel on the left whereas BP2 belongs to the panel on the right. We take 
$\sqrt{s}=13$ TeV.
}
\label{fig::scan-delta}
\end{figure}

Taking all the constraints without assuming that the model provides a dark matter candidate we have evaluated the deviation $\delta$, in all three inclusive cross sections, by scanning over a large region of the IDM parameters. 
Our results are displayed in Fig.~\ref{fig::scan-delta} for two large classes.
\begin{itemize}
\item {\bf A:}  In this category, we consider models with exact degeneracy in the scalar sector: $M_X=M_A=M_{H^\pm}$ and scan over $M_A$, with $100<M_A<1000$ GeV. 
In this case, we have $\lambda_L=\lambda_3=\lambda_A$. 
We scan over $0<\lambda_L<8$.

\item {\bf B:} In this category, to which BP2 belongs, $M_A=M_{H^\pm}$, $\Delta M =M_A-M_X = 100$ GeV is kept constant. 
The scan over $M_X$ is as previously done. To maintain uniformity with the display of the previous scan, we take $M_A$ as a variable so that the two scans show the same range $M_A$.
In this category $\lambda_A=\lambda_3$ and $\lambda_A>\lambda_L$, the difference between $\lambda_A$ and $\lambda_L$  increases as $M_X$ increases. 
This means that for the same $\lambda_L$ in models {\bf A}, $\lambda_{A,3}$ (couplings to $hAA, h H^{\pm}H^\mp$) can be larger, and hence we would expect larger deviations. 
\end{itemize}
The most important observation from Fig.~\ref{fig::scan-delta} is that deviations larger than $|\delta|=2\%$ (above the theoretical uncertainty we discussed for the SM predictions) are possible in a non-negligible part of the parameter space. 
The largest deviations occur for $(400<M_X<500)$ GeV; this maximum deviation can be understood on the basis that larger masses make the couplings larger, but for too large masses, some decoupling takes place. As expected, the largest deviations occur with models {\bf B} where these deviations can  reach   -4\% (at $M_X=480$ GeV for $Zh$). For models of type {\bf A} (the all-degenerate case), the discrepancy is more modest with the largest deviations from the SM reach -3\%.

Observe that a characteristic of the IDM is that $\delta$ is negative throughout the parameter space and is approximately the same in the three processes. 
Another important observation is that the largest deviations correspond to the largest values of $\lambda_L$, $6 <\lambda_L<8$. Deviations may be even larger for models like BP0 with a small $M_X$, not represented in Fig.~\ref{fig::scan-delta}. 
For BP0 the deviations from the SM reach almost $-6\%$ in all 3 channels, $-5.82\%, -5.76\%,-5.76\%$ respectively in $Zh, W^-h, W^+h$. \\

The benchmark points we have selected and the two large regions {\bf A} and {\bf B} reveal that deviations greater than the projected theoretical uncertainty ($2 \%$) are possible. Unfortunately, current experimental data have not reached the precision required to probe the IDM indirectly in the $Vh$ channels. The signal strength in the dominant channel $h \to b \bar b$ is at present only $25\%$ of the SM value\cite{ATLAS:2018kot,CMS:2018nsn}. But interest in other channels is growing (see for example $\tau \tau$ \cite{ATLAS:2023qpu}). New analyses within the high luminosity LHC (HL-LHC) have been conducted. In the $b \bar b$ $Zh$ channel alone, the experimental precision is predicted to reach $5\%$~\cite{ATLAS:2018jlh,CMS:2018qgz}. Further analyses of the different $Zh$ channels will certainly improve so that their combinations could provide an experimental uncertainty of $~2\%$. The improvement in LHC data, not just for $Vh$,  will also help achieve a better extraction of the PDFs which in turn will improve the uncertainty (PDF) of the theoretical prediction of $Vh$. With an overall (experimental/theoretical) uncertainty of about $3\%$, the $Vh$ channel will, in the HL-LH era, qualify as a precision channel that will be sensitive to the indirect effects of the IDM.  The HL-LHC would certainly be sensitive, through inclusive cross section analyses of these channels, to the effect of some parameter space of the IDM, in particular some of those covered by the scan of type B or the benchmark point of type BP0. With the tools we have developed, investigations of other regions of the parameter space than the ones we studied here are worth scrutinising. 

\begin{figure}[H]
\centering
\includegraphics[width=\textwidth,height=0.65\textwidth]{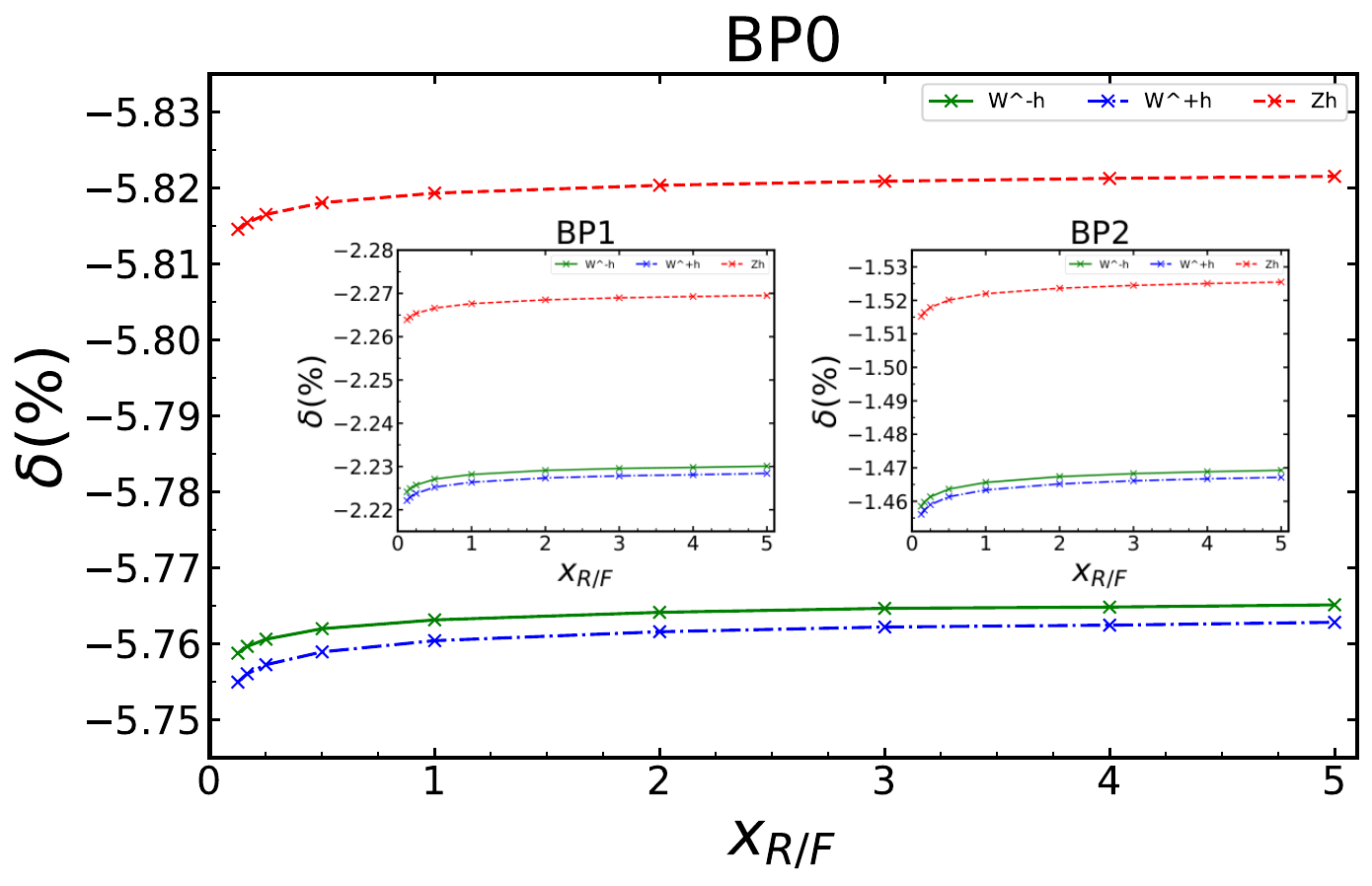}
\caption{The $\mu_{R/F}$ dependence of the normalised relative correction  $\delta$  for benchmark point  BP0  for the three Higgs-strahlung processes. The $\mu_{R/F}$ dependence is tracked through the variable  $x_{R/F}$, see Eqt.~\ref{renox}. The inserts correspond to BP1 (left) and BP2 (right).}
\label{fig::mudelta}
\end{figure}

As expected, the normalised ratio $\delta$ is practically insensitive to the factorisation / renormalisation scale $\mu_{R/F}$ in the three channels and for all three benchmarks, as Fig.~\ref{fig::mudelta} shows. Therefore, the predictions about the deviation from the SM are robust.

\begin{figure}[H]
\centering
\includegraphics[width=0.9\textwidth,height=0.54\textwidth]{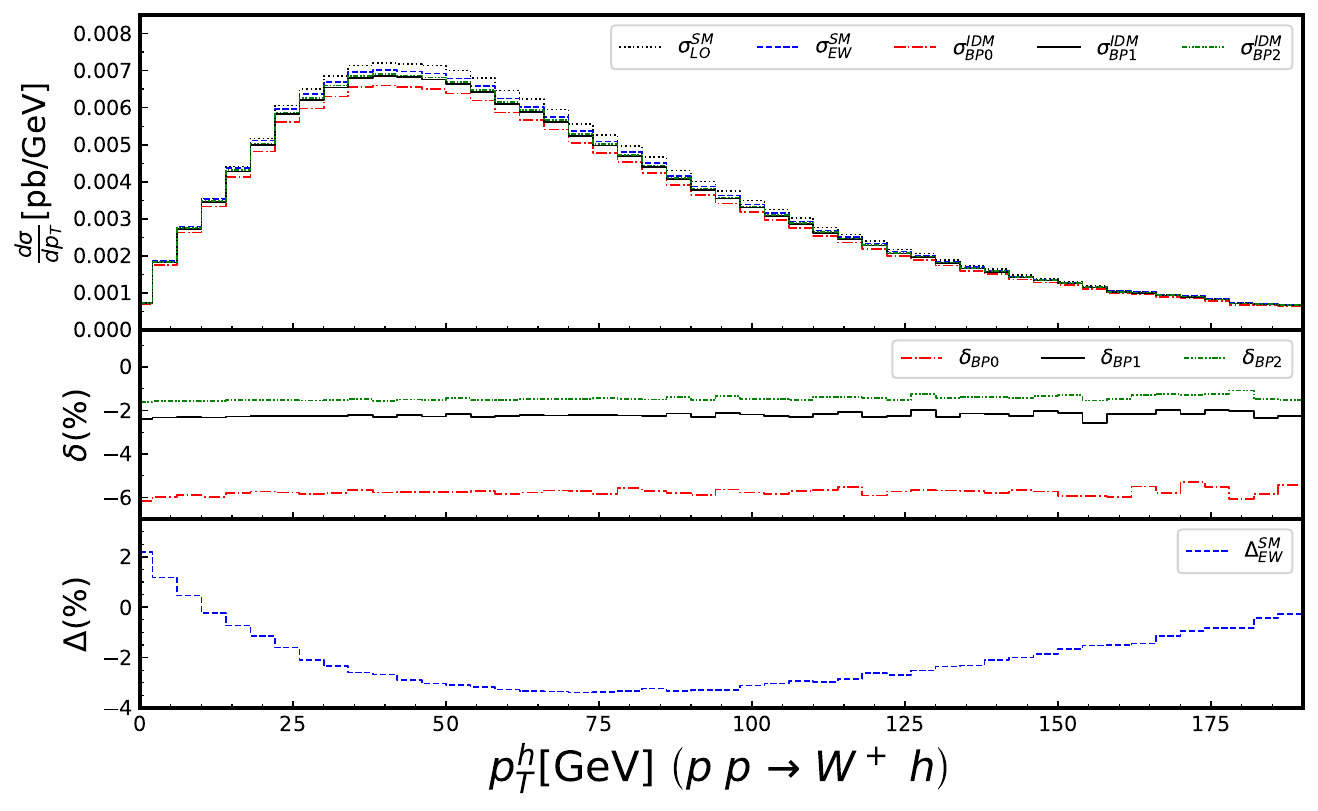}
\includegraphics[width=0.9\textwidth,height=0.54\textwidth]{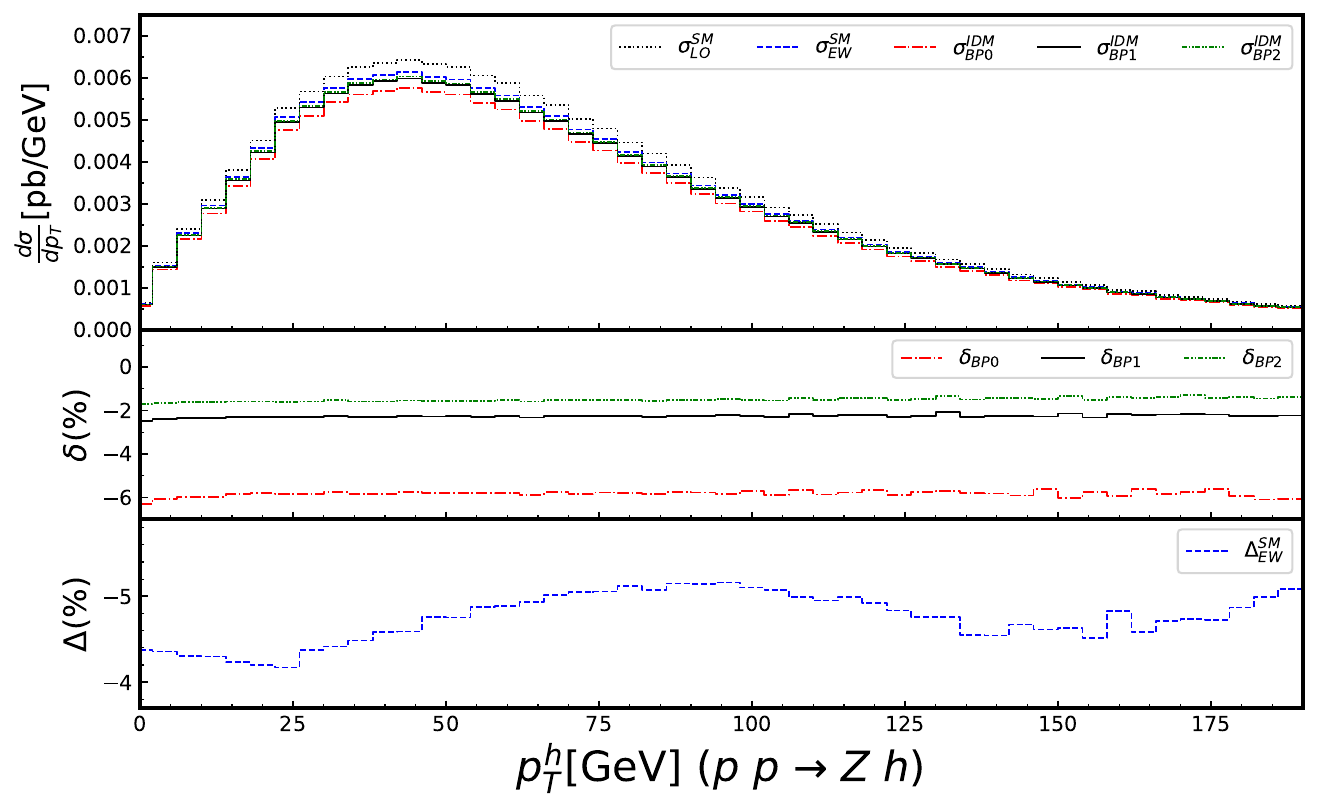}
\caption{We take two illustrative examples ($W^+h$ and $Zh$ channels) to show the impact of the new scalars on the $p_T^h$ distribution. The LO(black dots line), SM EW(blue dashed line) and IDM EW(red dots  and dashed line for BP0, black line for BP1 and green dots  and dashed for BP2) at 13TeV LHC. The percentage shift $\delta$ for the distributions is shown in the panel below each distribution. In the panel on the bottom we show the percentage correction for the SM electroweak correction, $\Delta\equiv \Delta_{EW}^{{\rm SM}}=(\frac{{\rm d} \sigma_{EW}^{SM}}{{\rm d}p_T^h}-\frac{{\rm d} \sigma_{LO}^{SM}}{{\rm d}p_T^h})/\frac{{\rm d} \sigma_{LO}^{SM}}{{\rm d}p_T^h}$ .}
\label{fig::ptdisidm}
\end{figure}

We have also analysed whether the shifts from the SM in the total cross sections manifest themselves through characteristic distributions that would further help in distinguishing the IDM. We have looked at the transverse momentum and rapidity distributions of the Higgs for the three reference points BP0, BP1 and BP2. We find that, independently which of the three cross sections or the benchmark point is considered,  the shifts in the cross sections are practically equally distributed particularly for $p_T^{h}$, while there is some distinctive structure for the $y^h$ distribution from the electroweak correction within the SM, see Fig.~\ref{fig::ptdisidm}.  

\begin{figure}[!htbp]
\centering
\setlength{\abovecaptionskip}{-0.3cm}
\includegraphics[width=\textwidth,height=0.6\textwidth]{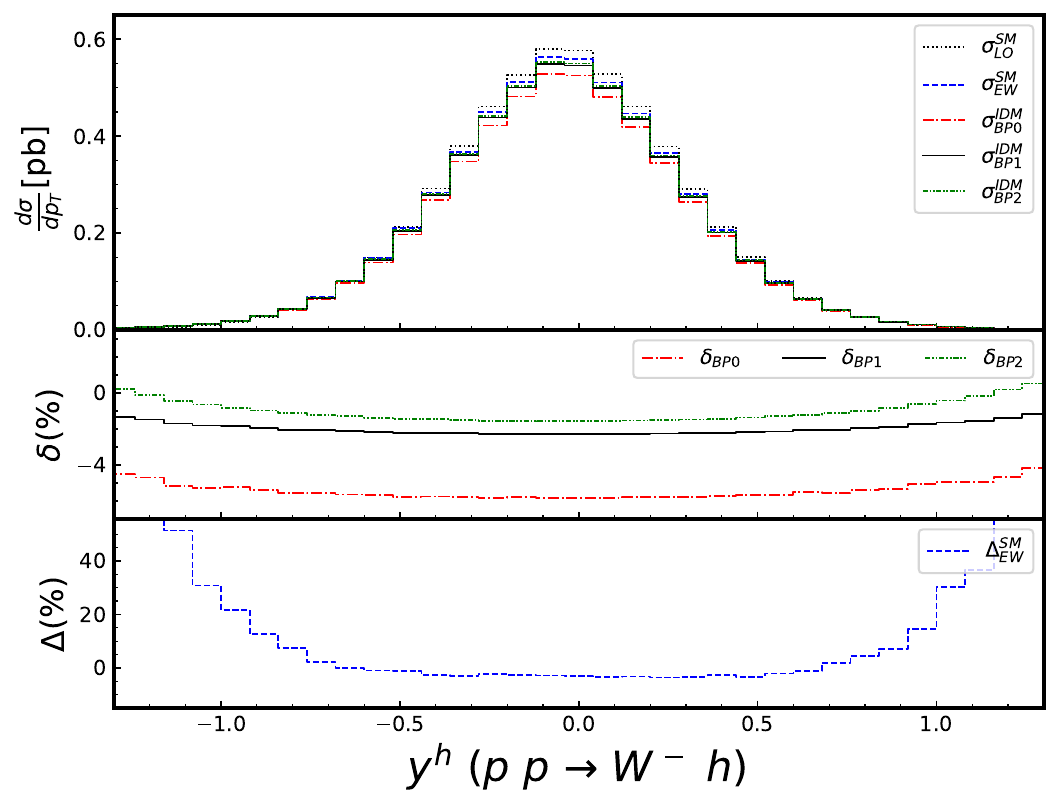}
\includegraphics[width=\textwidth,height=0.6\textwidth]{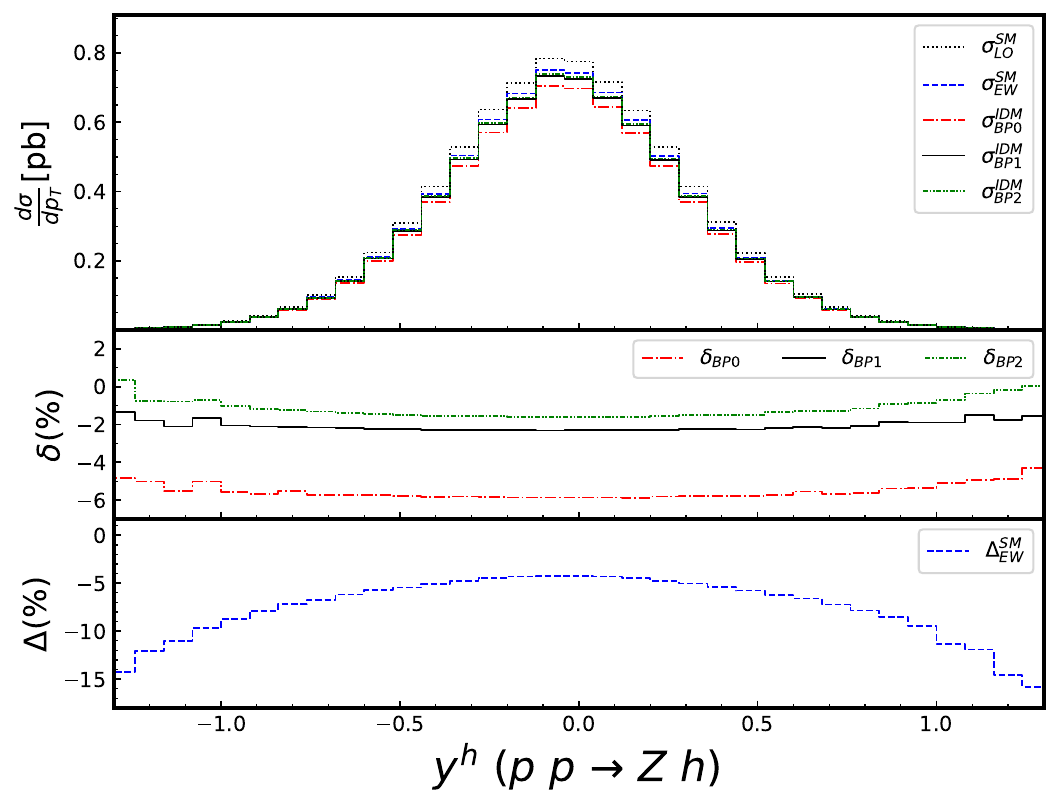}
\caption{As in Fig.~\ref{fig::ptdisidm} but for the rapidity distribution for the $W^-h$ and $Zh$ channels.}
\label{fig::ydisidm}
\end{figure}

For the rapidity distribution, $y^{h}$, the electroweak SM corrections are revealed through a quite distinctive feature, both for $W^\pm h$ and $Zh$,  particularly around large $|y^h|$. The added effect of the IDM  is not as subtle as on the $p_T^h$ distribution with a small disturbance from the SM results that affects the central events around $y^h \sim 0$. However, note that these are the regions where the cross sections are the smallest; see Fig.~\ref{fig::ydisidm}. The effect of the IDM may therefore not be resolved.

\subsection{Scenarios with DM constraints}
\label{sec:hv_idm_dm}
We now impose the DM constraints. As noted earlier, direct detection requires extremely small values for $\lambda_L$, $\lambda_L$ of the order of percent or smaller. For not too heavy spectrum and to conform to custodial symmetry, the other couplings $\lambda_3,\lambda_A$ are smaller than 1, see Sect.~\ref{sec::currentconstraints}. The results of the scan
in Fig.~\ref{fig::scan-delta} give us an indication of the expected deviation $\delta$ in these scans. The deviations for small $\lambda_L< 1$ are extremely small, $\ | \delta\ |<0.5 \%$ (see Fig.~\ref{fig::scan-delta}). This is confirmed by the analysis of the benchmark points BP3-BP5 as Table~\ref{tab::resultsII13} shows (the results are essentially the same for the LHC at 14 TeV). The deviations when the DM constraints are taken into account are at best at the per-mil level, this is in $W^\pm h$ $\delta$ for BP4 (this is understood on the basis that this BP has the largest coupling $\lambda_3=\lambda_{hAA}$ combined with the smallest value of the neutral scalar $A$). Even in this most favourable situation the deviation is below $0.4 \%$. When DM constraints come into play, the effect of the IDM in precision indirect effects of $pp \to Vh$  cross sections at the LHC (including its high luminosity upgrade)  are not observable.

\renewcommand{\dblfloatpagefraction}{1}
\begin{table}[!htb]
\renewcommand\arraystretch{1.5}
\begin{center}
\begin{tabular}{| p{1.8cm}<{\centering} |p{2.2cm}<{\centering} |p{2.2cm}<{\centering}| p{2.2cm}<{\centering}|}
\cline{2-4}
\multicolumn{1}{c|}{}& BP3    & BP4 & BP5\\
\hline
$Zh$          & 0.07   &  $\mathcal{O}(10^{-3})$    &   -$\mathcal{O}(10^{-3})$       \\
\hline
$W^-h$        & 0.20   &  0.35      &   -$\mathcal{O}(10^{-3})$    \\
\hline
$W^+h$        & 0.19   &  0.34      &  $ - \mathcal{O}(10^{-3})$    \\
\hline
\end{tabular}
\end{center}
\caption{The deviation $\delta$ in $\%$ for the benchmark points BP3, BP4 and BP5 (with dark matter constraints imposed) on the $Vh$ cross sections at the 13 TeV LHC.}
\label{tab::resultsII13}
\end{table}

\section{Summary}
\label{sec::summary}
Higgs-strahlung processes at the LHC are entering the era of precision calculations. Computations at N$^3$LO QCD are among the latest achievements~\cite{Baglio:2022wzu}. Concurrently, the accuracy of the PDF sets have greatly improved, making these processes prime observables for investigating indirect, virtual effects of New Physics. 
In this study, we considered the effects of the Inert Doublet Model (IDM) as a prototype for a more general class of models to be investigated through their indirect effects at the LHC. While the indirect effects of the IDM are fundamentally of electroweak origin, we also revisited the calculation of the NLO QCD and EW corrections within the SM. IDM apart,  one of our findings indicates that the $\gamma q \to W h q_j$ production, which is part of the NLO $W^\pm h$ cross section, can be non-negligible in the calculation of the inclusive cross sections. A  $p_T^W$ cut can, however, reduce this contribution.

A key objective of this study, which is part of a broader effort, is to develop a general tool to perform one-loop calculations at the LHC in and beyond SM theories. This tool aims to address both NLO QCD and one-loop electroweak corrections, incorporating scale and PDF uncertainties in SM calculations while precisely computing the effects of New Physics models within the same framework. This study, which showcases the IDM, furnishes a first excellent testing ground for such investigations and a fuller exploitation of the LHC. \\
The main motivation for this study was also to determine whether existing  constraints, such as Electroweak Precision Observables (EWPO), non-observation at the colliders of the model's physical states, and limits on the partial widths of the Higgs, still allow for the potential future observation of deviations in processes like $W^\pm h$ and $Zh$ at the LHC. Our results suggest that deviations in the total yield are indeed possible, even after accounting for PDF and parametric uncertainties of the SM cross sections. We have investigated these theoretical uncertainties at some length.  
Indeed, an important question is to first investigate whether the additional New Physics effects can be large enough to stand above the uncertainties (scale and PDF) of the precision calculation of the SM contributions before addressing the possibility whether such channels could be exploited for the indirect effects of New Physics at the LHC, in particular in its future high luminosity upgrade. \\

Generally, deviations are observed in both the charged and neutral channels. However, no clear discernible structure is found in the distributions ($p_T^h, y_h$) we examined.
We find that while \underline{current} experimental data from the LHC on $Vh$ production are not yet precise enough (compared to theoretical uncertainties), future data from a high-luminosity LHC could be exploited to restrain the parameter space of the model. In particular, experimentalists have recently shown renewed interest in $Vh$ production in the HL-LHC set-up, particularly with $h \to b\bar{b}$ channel. For example, in preliminary analyses of this channel a precision of $5\%$ is already foreseen for $Zh$ production \cite{ATLAS:2018jlh,CMS:2018qgz} in the high luminosity LHC runs.
We are confident that by the time the HL-LHC data are exploited, other $Vh$ channels will have been included and further improvements in the $h \to b \bar b, h \to \tau \bar{\tau}$ signature will have been achieved. 
At the same time, the accuracy of the PDF sets will have improved through the exploitation of large LHC data sets. This will reduce an important part of the theoretical uncertainties. Combinations of different channels with $V=W^\pm$ and $Z$ will reduce the experimental accuracy to a better level than the actual projections for HL-LHC with respect to the measurements of the $Vh$ cross sections\cite{ATLAS:2018jlh,CMS:2018qgz}.

At the same time, the accuracy of the PDF sets will have improved through the exploitation of more data that will have collected and analysed by then. This will reduce an important part of the theoretical uncertainties. The latter will benefit from the inclusion of much refined,  future ($N^{n\geq 3}$LO, particularly QCD) computations. In such an optimistic situation, the indirect effects of the IDM will be sensitive to a larger parameter space of the model than the ones we uncovered in the present study.
These developments stress the need for tools like the one we are developing for handling EW  corrections at the LHC and the impact of virtual corrections from heavy physics at the LHC.

When the IDM is considered as a model providing a possible dark matter candidate and constraints from the dark matter sector, particularly direct detection, are imposed, the model becomes highly constrained. Under these conditions, we have found that deviations in precision measurements in all Higgs-strahlung processes become too minuscule to stand out above the theoretical SM accuracies (present and probably future), let alone the expected experimental precision that is foreseen in future LHC data on $Vh$ production.

\renewcommand{\thesection}{\Alph{section}}
\setcounter{section}{0}


\section{Appendix: Recasts and reinterpretation of LHC cross sections}

Due to the $Z_2$ symmetry, the new scalars can only be produced in pairs through, predominantly, Drell-Yan-like processes at the LHC. The most important signatures \cite{Belanger:2015kga,Datta:2016nfz,Tsai:2019eqi} consist of large missing transverse energy associated with a certain number, $l$, of leptons (as decay products of the $W/Z$ vector bosons either on-shell or off-shell)
\begin{eqnarray}
\label{mono-lep}
\label{eq:l1}
H^\pm X &\to & W^\pm XX \to l^\pm + \s{E}_T, \qquad l=1 
\label{eq:l2} \\
A X &\to & (Z) XX \to l^+ l^- + \s{E}_T, \qquad l=2 \quad {\scriptstyle (\text{C1N2-WZ, mono-Z})} \\
\label{eq:l3}
H^\pm A &\to &  (W^\pm) (Z) XX \to l^{\prime \pm} l^+ l^- + \s{E}_T, \qquad l=3 \\
\label{eq:ll21}
H^+ H^- &\to &  (W^+) (W^+) XX \to l^{+}  l^{\prime -}  + \s{E}_T, \qquad l=2. \quad {\scriptstyle (\text{C1C1-WW})} 
\label{eq:l2l2}
\end{eqnarray}

The possibility of fully hadronic decays of the vector bosons has also been considered recently~\cite{ATLAS:2021yqv}.  
These dominant processes are gauge-mediated. They are perfectly predicted when the mass spectrum is provided. 
It is important to stress that these final-state topologies are exactly the same as those that ensue from the simplified electroweakino models, winos/higgsinos (with all other superparticles too heavy,  hence decoupled)\footnote{See for example the wiki page of the ATLAS SUSY Searches:\\ {\tt https://twiki.cern.ch/twiki/bin/view/LHCPhysics/SUSYCrossSections}}.
The categorisation C1C1-WW, C1N2-WZ is borrowed from the ATLAS analyses. 
In this case, the identification with the IDM can be made with $X \to \chi_1^0, H^\pm \to \chi_1^\pm, A \to \chi_2^0$.  
These channels were used to check the viability of our IDM models (which constitute the large parameter scan of Fig.~\ref{fig::scan-delta} as well as the benchmark points BP1, BP2, and BP5. 
To avoid the strong constraint of the parameters $T$, for the large parameter scan, we have used two classes: {\it i) } $M_X=M_A=M_{H^\pm}$ and {\it ii) } $M_A=M_{H^\pm}=M_X+100$ GeV with $100<M_X<1000$ GeV, $0<\lambda_L<8$ (Fig.~20). 
The first case trivially passes the LHC constraint as one can not trigger on any visible track. 
For class {\it ii)}  the topologies listed above can be fully exploited, the reinterpretation of the wino exclusion zone (in~\cite{ATLAS:2019lff}) is then made taking into account the fact that the branching ratios $H^{\pm} \to W^\pm X$ and $A \to XZ$ are both $100\%$ and that both $W$ and $Z$ in these decays are produced \underline{on-shell}. 
Translating the electroweakinos results of the ATLAS analyses \cite{ATLAS:2019lff} to the IDM shows that the regions with $M_X> 100$ GeV and $M_A=M_{H^\pm}=M_X+100$ GeV are allowed. 

This reinterpretation is fully warranted. First (as shown in Table.~2 of~\cite{Banerjee:2021oxc} for low masses, but a trend that we have verified to be valid for all masses) the cross sections for the production of the {\bf IDM scalars} are at least an order of magnitude smaller than the production of (the same mass) {\bf winos / higgsinos (fermions)} that lead to the same topologies.  
One might argue that the spin enhancement in the SUSY case over the IDM may be offset by the choice of cuts that would be more beneficial for fermions than scalars({\it i.e. reducing the cross sections for the production of scalars}). 
First, the choice of cuts on the final decay products is designed to decrease the SM backgrounds but,  taking as an example \cite{ATLAS:2019lff} for CC1-WW,  the analysis optimises the signal-to-background separation using the $M_{T2}$ variable divided in multiple bins and does not use a variable that is {\em spin-sensitive} such as one that is explicitly dependent on the angular separation of the final-state particles. 
Second, the optimal experimental cuts can not reduce the ratio by more than an order of magnitude between the yield in SUSY and the yield in the IDM. 
Therefore, any point that is allowed in the SUSY case would certainly pass the LHC constraint in the case of the IDM.

The topologies with fully hadronic final states and missing energy as studied, for example, in~\cite{ATLAS:2021yqv} turn out not to be relevant for the scenarios we have considered ($M_X> 100$ GeV, $M_H^\pm=M_A=M_X+100$ GeV or, of course, the fully degenerate case).  

For the low-spectrum benchmark points with $M_X <M_W$,
it should first be noted that BP3, BP4, BP0 correspond to (respectively) the benchmark P59, P60, and D in \cite{Banerjee:2021oxc}. 
Although a reinterpretation as done above for the other points would have proved sufficient, we have here performed a recast  within the IDM with the help of {\tt MadAnalysis 5}~\cite{Conte:2012fm,Conte:2014zja,Dumont:2014tja,Conte:2018vmg,Araz:2020lnp}, exploiting 
\begin{itemize}
\item $2\ell+\slashed{E}_T$~\cite{ATLAS:2019lff,Araz:2020dlf} 
from electroweakinos/slepton pair production  
\item $2\ell + \slashed{E}_T, 3\ell + \slashed{E}_T$~\cite{CMS:2017moi,validation:electroweakinos} from the chargino-neutralino pair production, 
\item $2\ell+\slashed{E}_T$~\cite{CMS:2017ret,validation:monoZ}  mono-$Z$ searches 
\item $\ell + \slashed{E}_T$~\cite{ATLAS:2019lsy,validation:Wprime}
$W^\prime$ searches. 
\end{itemize}
This recast confirms the result that would have been based solely on the simplified reinterpretation using the first channel from electroweakino production and lends credence to the exploitation of the reinterpretation. For more details, we urge the interested reader to consult \cite{Banerjee:2021oxc}.

The IDM can also lead to monojet final states ($XXj,XAj$. $j$ is a jet). These signatures can be produced through $\{gg,qq \} \to h g$ or $hq$ with decay $h \to  \chi \chi,\ \chi=X,A$. These processes are relevant only when $\lambda_L$ is not too small ($XXj$) or the mass splitting between $A$ and $X$ is small. Such manifestations are not relevant for our scenarios. The interested reader may want to consult \cite{Dercks:2018wch}. 

With the tools we have at our disposal, we can now entertain investigating a larger array of parameter space of the IDM providing cross sections for the IDM at NLO and performing dedicated optimised recasts. However, for the parameter space chosen in this paper, models {\bf A} and models {\bf B} (the scans), which feature total mass degeneracy or small $\Delta M$, complete recasts incorporating present and future Run 2 data will have little to no impact compared to an interpretation as suggested in this Appendix.

~\\
~\\
~\\

\acknowledgments
H.S. is supported by the National Natural Science Foundation of China under Grant No.~12075043. Y.Z. is supported by the National Natural Science Foundation of China under Grant No. 12475106 and the Fundamental Research Funds for Central Universities under Grant No. JZ2023HGTB0222. The authors thank Jean-Philippe Guillet for discussions and careful reading of the manuscript.

~\\
~\\
~\\

\bibliographystyle{JHEP}
\bibliography{refv2.bib}


\end{document}